# Role of topological surface states and mirror symmetry in topological crystalline insulator SnTe as an efficient electrocatalyst

*Qing Qu[1, 3†], Bin Liu[2†], Hongtao Liu[2], Jing Liang[2,3], Jiannong Wang[2,3], Ding Pan[2,4*] and Iam Keong Sou[1,2,3*]*

[1]Nano Science and Technology Program, The Hong Kong University of Science and Technology, Hong Kong, China.

[2]Department of Physics, The Hong Kong University of Science and Technology, Clear Water Bay, Hong Kong, China.

[3]William Mong Institute of Nano Science and Technology, The Hong Kong University of Science and Technology, Hong Kong, China

[4]Department of Chemistry, The Hong Kong University of Science and Technology, Hong Kong, China.

[†] These authors contributed equally to this work.

* Corresponding authors

ABSTRACT：The surface orientation dependence on the hydrogen evolution reaction (HER) performance of topological crystalline insulator (TCI) SnTe thin films is studied. Their intrinsic activities are determined by linear sweep voltammetry and cyclic voltammetry measurements. It is found that SnTe (001) and (111) surfaces exhibit intrinsic activities significantly larger than the (211) surface. Density functional theory calculations reveal that pure (001) and (111) surfaces are not good electrocatalysts, while those with Sn vacancies or partially oxidized surfaces, with the latter as evidenced by X-ray photoelectron spectroscopy, have high activity. The calculated overall performance of the (001) and (111) surfaces with robust topological surface states (TSSs) is better than that of the lowly symmetric (211) surface with fragile or without TSSs, which is further supported by their measured weak antilocalization strength. The high HER activity of SnTe (001) and (111) is attributed to the enhanced charge transfer between H atoms and TSSs. We also address the effect of possible surface facets and the contrast of the HER activity of the available active sites among the three samples. Our study demonstrates that the TSSs and mirror symmetry of TCIs expedite their HER activity.

## Introduction

The limitation of fossil-fuel-based energy resources and the increase of carbon-dioxide-related pollution have amplified the urgency for exploiting carbonless energy.[1] Among carbon-free energy resources, hydrogen ($H_2$), is particularly popular because it contributes no environmental pollutants.[2] Currently, electrochemical water splitting is regarded as the most promising eco-friendly and economical method for sustainable $H_2$ production.[3,4] Platinum-group metals are currently the most widely used electrocatalysts for hydrogen evolution reaction (HER), which however are expensive and scarce metals that are also poisoned easily by organic functional group and inorganic anions.[5-7] Accordingly, enormous research effort is currently being directed towards the development of low-cost and highly efficient HER electrocatalysts based on non-noble-metal key active materials.[2, 8-11]

Recently, studies on a number of topological materials were reported both experimentally and theoretically to provide a novel direction for the design of high-efficiency non-noble-metal catalysts for HER.[12-16] The existence of nontrivial topological surface states (TSSs) in these materials is considered to play an important role in catalytic processes. Particularly, such nontrivial TSSs in topological insulators (TIs) are the results of band inversion between the bulk conduction and valence bands. The time-reversal symmetry (TRS) and particle number conservation protect the TSSs from backscattering by nonmagnetic impurities.[17] In general, TSSs are robust and are protected against surface modifications.[18] Thus, local perturbations such as light doping, defect creation, chemisorption of air molecules, or even slight oxidation under exposure to the ambient environments would not annihilate the TSSs.[16,19-22] It has been well accepted that the highly desirable functionality in TI-based HER electrocatalysts is attributed to the TSSs as revealed by first-principles density functional theory (DFT) calculations.[15,23,24] In the past few years, several bismuth chalcogenide-based TIs have been proved to be promising candidates of photocatalysts,[13] and $Bi_2Te_3$ thin films with partially oxidized structures or Te vacancies[16] were shown to display high HER activities, providing evidence of the important role of TSSs in catalytic processes.

In 2011, Fu has extended the family of TIs by introducing the topological crystalline insulators (TCIs) [25,26] where the topology is protected by a point-group symmetry of the crystal lattice rather than by TRS. The first theoretically predicted and experimentally realized TCI materials are IV-VI semiconductors, with SnTe as a representative,[25-27] which has metallic surface states with an even number of Dirac cones on high-symmetry crystal surfaces such as {001}, {110} and {111}. These TSSs form a new type of high-mobility chiral electron gas, which is robust against disorder and topologically protected by reflection symmetry of the crystal with respect to the {110} mirror plane.

To the best of our knowledge, research on TCI thin film materials directly used as electrocatalysts in HER with efficient performance has not been reported so far. It has been reported that there are two types of TCI surface states with qualitatively different electronic properties depending on the surface orientation:[28] type-I TCI surface states on the (111) surface have the properties that their Dirac points are located at the time-reversal-invariant momenta (TRIMs), while the locations of Dirac points for type-II TCI surface states on the (001) and (110) surfaces are deviated from TRIMs. More importantly, the interplay between topology and crystal symmetry is the key character of TCIs, which represents that if the mirror symmetry of SnTe is broken, the gapless surface states will be gapped.[29] For example, the lattice distortion and mirror symmetry breaking induced Dirac gap opening in TCI $Pb_{1-x}Sn_xSe$ was revealed by scanning tunneling microscopy[30,31] and angle-resolved photoemission spectroscopy,[32,33] which is consistent with theoretical calculations.[34] Wei *et al* reported that the broken mirror symmetry due to the cubic-rhombohedral structural phase transition can tune topological transport in PbTe/SnTe heterostructures.[35] DFT calculations carried out by Dagdeviren *et al* show that local symmetry breaking obstructs the TSSs of stepped (103) surface of SnTe by a threefold reduction of its spectral weight as compared with the ideal (001) surface.[36]

In this work, we prepared SnTe (111), SnTe (001) and SnTe (211) TCI thin films by molecular beam epitaxy (MBE) for studying the role of TSSs and mirror symmetry in electrocatalytic HER performance. SnTe (111), SnTe (001) and SnTe (211) were chosen as the representatives for type-I, type-II surfaces, and a low-mirror-symmetric surface, respectively. It is revealed that the SnTe (001) and SnTe (111) samples exhibit lower overpotentials, Tafel slopes and charge-transfer resistances than those of the SnTe (211) sample. More importantly, it is well-known that the surface morphology influences significantly the electrocatalytic activities, and thus per-site turnover frequencies (TOFs) of these samples were extracted to determine the intrinsic activities of these surfaces. It was found that the TOF values of the SnTe (111) and SnTe (001) samples are comparable but much larger than that of the SnTe (211) sample. Our DFT calculations show that the Sn vacancy or the surface partial oxidation lifts the Dirac cones and promotes the charge transfer from the H atoms to the surfaces, which makes the H adsorption easier, so the HER performance is enhanced. The findings from the X-ray photoelectron spectroscopy and magneto-transport measurements performed on these samples are consistent with the theoretical calculations. We have also addressed the structural differences among the three orientations regarding the existence of various facets at the surfaces. Our observations indicate that even though the (211) sample has the highest number of active sites attributed to its roughest surface, however, they are much less active and overwhelmed by the TSSs effect that is only available and robust for the samples with (001) and (111) orientations.

## Results

A seemingly straight forward approach to fabricate the three SnTe thin films with different orientation required by this study is to grow SnTe (111), SnTe (001) and SnTe (211) directly on GaAs (111)B, GaAs (001) and GaAs (211)B substrates. However, this approach was found to be unachievable for the reasons to be addressed below.

The high-resolution X-ray diffraction (HRXRD) profile of a SnTe thin film directly grown on a GaAs (111)B substrate is shown in Figure S1 in supplementary information, where one can see that the resulted SnTe thin film is oriented along the [001] direction. Taskin *et al* have previously demonstrated that SnTe (111) can be grown directly on $Bi_2Te_3$ (001) surface,[37] which was attributed to the small lattice mismatch of 1.5% between these two surfaces and the similarity of their surface unit cells. In the current study, in order to avoid the effects of the TSSs of $Bi_2Te_3$, we have adopted the following approach for fabricating a SnTe (111) layer. A ZnSe (111) buffer layer was firstly deposited on a GaAs (111)B substrate followed by the growth of a $Bi_2Te_3$:Fe (001) layer. Our previous study reveals that light doping of Fe in $Bi_2Te_3$ can destroy the TSSs of $Bi_2Te_3$ as supported by the quenching of the weak antilocalization (WAL) signature.[16] The growth of the top SnTe (111) layer was then followed to complete the growth of this sample.

On the other hand, it was found that SnTe (001) thin film can be directly grown on a GaAs (001) substrate. For the growth of the SnTe (211) thin film, since the GaAs (211)B substrate is not an epi-ready substrate, the required wet chemical etching process was found to generate a rather rough surface, thus a thick ZnSe buffer layer was firstly grown prior to the growth of the top SnTe (211) layer.

**Structural Characterizations of MBE-Grown SnTe (111), (001) and (211) Samples.**
Figure S2 in supplementary information shows the sample structures of the MBE-grown SnTe (111), SnTe (001), and SnTe (211) samples and the reflection high-energy electron diffraction (RHEED) patterns captured during the growth. Figure S2a displays the sample structure of the SnTe (111) sample. Figures S2b, c exhibit the RHEED patterns of the $Bi_2Te_3$: Fe (0001) layer and the top SnTe (111) layer, respectively, where the 4-index notation is used in RHEED pattern analysis of $Bi_2Te_3$. Figure S2b displays the streaky RHEED patterns (corresponding to the reciprocal lattice) of hexagonal $Bi_2Te_3$:Fe (0001) surface when the incident electron beam is along the $[10\bar{1}0]$ (left) and $[2\bar{1}\bar{1}0]$ (right) direction and the spacing ratio of the streaks of the two azimuths is $\frac{1}{\sqrt{3}}$, which is attributed to the $\sqrt{3}$ ratio in the real space lattices shown at the bottom. Figure S2c shows the streaky RHEED patterns of the triangular SnTe (111) surface when the incident electron beam is along the $[2\bar{1}\bar{1}]$ (left) and $[10\bar{1}]$ (right) direction, and the spacing ratio of the streaks of the two azimuths is also $\frac{1}{\sqrt{3}}$, which is attributed to the $\sqrt{3}$ ratio in the real space lattices shown at the bottom. The sample structure of the SnTe (001) sample is shown in Figure S2d. Figure S2e displays the RHEED pattern of the SnTe layer when the incident electron beam is along the SnTe [100] (left) and [110] (right) azimuth. The dash lines below the RHEED pattern on the left corresponds to the surface reconstruction streaks. The left and right RHEED patterns together show that the as-grown SnTe (001) surface has a (2 × 1) surface reconstruction. The spacing ratio of the streaks of the two azimuths is $\frac{1}{\sqrt{2}}$, which is attributed to the $\sqrt{2}$ ratio in the real space lattices shown at the bottom. The sample structure of the SnTe (211) sample is shown in Figure S2f. One RHEED pattern of the SnTe (211) surface is shown in Figure S2g. A video of the RHEED patterns of this surface is provided in supplementary information to show its single crystalline nature.

**Figure 1**a-c display the HRXRD profiles of the SnTe (111), SnTe (001), and SnTe (211) samples. Figure 1d-g show the powder diffraction files (PDFs) of the four crystalline materials contained in these samples, where only the peaks oriented along the normal of the sample surface are extracted for comparison. Figure 1a displays the HRXRD profile of the SnTe (111) sample. As can be seen, all $Bi_2Te_3$ layer peaks can be indexed as the (00$l$) direction and their measured 2θ values give a $c$-lattice parameter of ~30.478 Å, which closely matches the PDF values of $Bi_2Te_3$ ($c$ = 30.483 Å). The characteristic diffraction peak locating at 50.04° matches well with the standard 2θ value of SnTe (222) giving a lattice constant of 6.309 Å, which has a deviation of only 0.29% from the standard value of 6.33 Å. The diffraction peaks of ZnSe (111) and ZnSe (222) are overlapped with the strong peaks of GaAs (111) and GaAs (222) in this broad-scan profile due to the very small lattice mismatch (0.27%) between ZnSe and GaAs.[16,38,39] Figure 1b displays the HRXRD profile of the SnTe (001) sample. Two characteristic diffraction peaks locating at 28.235° and 58.365° match with the standard 2θ values of SnTe (002) and (004), respectively. From these two peaks, the lattice constant of the SnTe (001) sample is determined to be 6.32 Å, which agrees well with the standard value from the PDF (6.33 Å), indicating that the SnTe (001) layer is almost fully relaxed. Figure 1c shows the HRXRD profile of the SnTe (211) sample. The characteristic diffraction peak locating at 72.55° matches with the standard $2\theta$ values of SnTe (422) giving the lattice constant to be 6.378 Å, which is also close to the standard value of 6.33 Å.

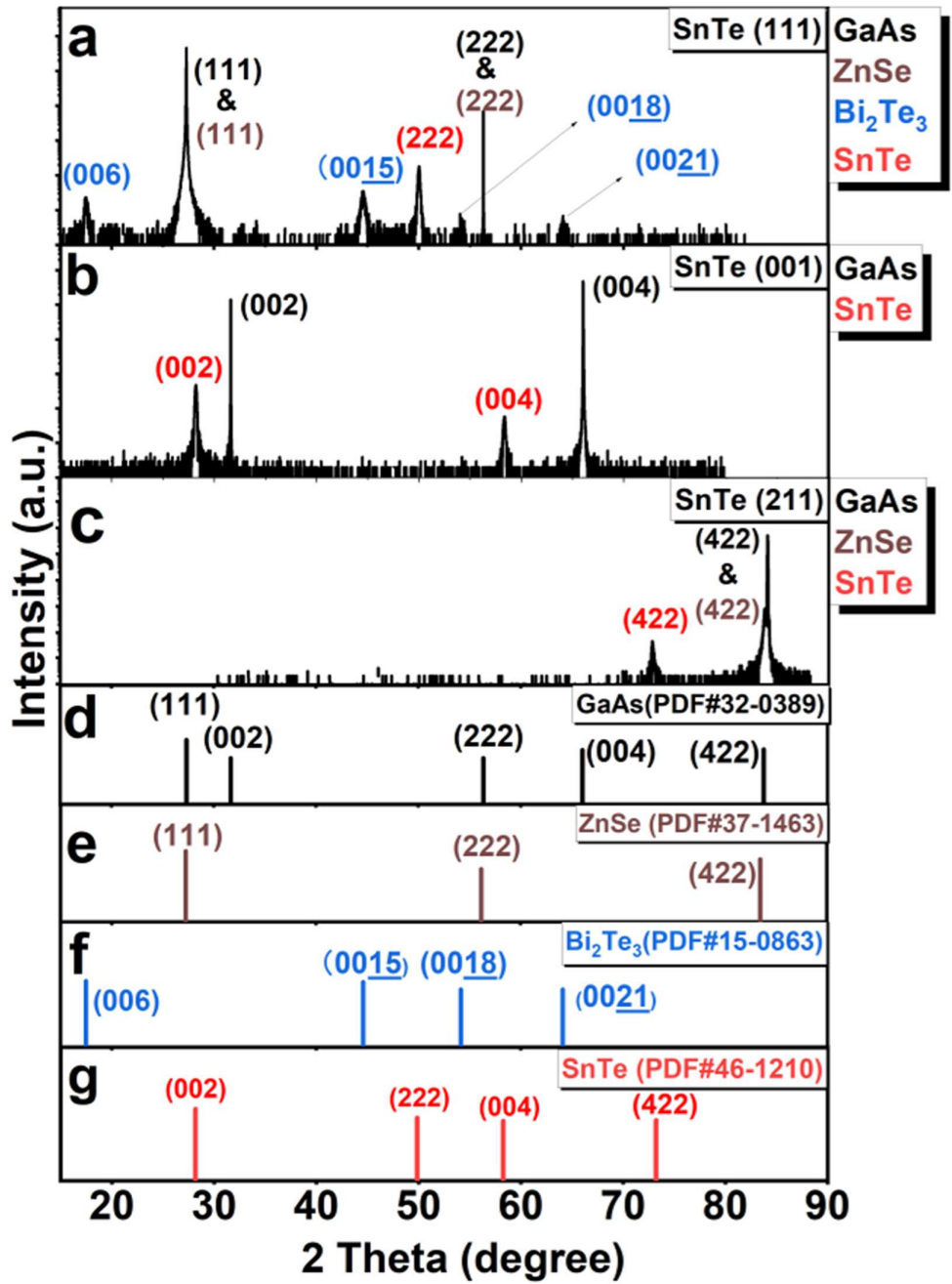

**Figure 1.** High-resolution X-ray diffraction (HRXRD) 2θ-ω scan profiles of the three MBE-grown thin film samples and the corresponding powder diffraction files (PDFs). (a) SnTe (111), (b) SnTe (001) and (c) SnTe (211) samples. (d–g) PDFs for the four materials contained in the three thin film samples.

The cross-sectional high-resolution transmission electron microscopy (HRTEM) images with fast Fourier transform (FFT) images and corresponding atomic arrangements of the three

SnTe samples studied in this work are shown in Figure S3. The thicknesses of the different layers of the three samples are determined from the cross-sectional TEM images shown in Figure S3a, g, m. As can be seen, the 73 nm SnTe (111) layer was grown on the 40 nm $Bi_2Te_3$:Fe (001) layer with a 5 nm ZnSe buffer layer (Figure S3a), the 70 nm SnTe (001) layer was grown on a GaAs (001) substrate directly (Figure S3g), and the 120 nm SnTe (211) layer was grown on the 630 nm (211) ZnSe buffer layer (Figure S3m). It should be noted that one should only take these values as the approximate nominal thicknesses since these samples have domains or the valley structures at their surfaces, which will be addressed later. It should also be mentioned that the cross-sectional TEM images of the three samples show that a thin oxidation layer of 5–8 nm exists at their surfaces, which is believed to be formed mainly via the reaction with the water and the G2 epoxy used during the TEM sample preparation.

The cross-sectional HRTEM image of the interface portion between the $Bi_2Te_3$: Fe (001) layer and the SnTe (111) layer of the SnTe (111) sample is displayed in Figure S3b. The energy dispersive spectroscopy (EDS) profile shown in Figure S4 indicates that the incorporated Fe concentration is about 0.8% (Table S1), which is close to the value of a similar $Bi_2Te_3$:Fe layer we reported previously.[16] The FFT pattern of the $Bi_2Te_3$:Fe (001) layer shown in Figure S3c reveals that two sets of hexagonal lattice appear in this thin film layer, indicating the existence of twin crystals,[16,40,41] attributed to the fact the (001)-oriented $Bi_2Te_3$ has a 3-fold symmetry, while the GaAs (111) substrate has a 6-fold symmetry. Figure S3d displays the schematic drawings of the two expected domains of the (001)-oriented $Bi_2Te_3$. Figure S3e, f display the FFT pattern and its corresponding atomic arrangement of the SnTe (111) layer. The FFT patterns shown in Figure S3c, e indicate the relationship of the $[111]_{SnTe}//$ $[001]_{Bi_2Te_3:Fe}$. Figure S3h, i, k, j, l show the cross-sectional HRTEM image, FFT patterns and the corresponding schematic drawings of the atomic arrangement of the (001)-oriented SnTe layer of the SnTe (001) sample. As shown in Figure S3h, the interface between the SnTe

(001) and GaAs (001) lattices as indicated by the arrow is not atomically flat, which is attributed to the outgassing of the GaAs (001) substrate without applying an As over pressure (which is not available in our MBE system) and the large lattice mismatch between SnTe and GaAs. Figure S3n exhibits the cross-sectional HRTEM image of the SnTe (211)/ZnSe (211) interface portion. It can be seen that the interface between the SnTe (211) and ZnSe (211) lattices as indicated by the arrow shows a wavy feature, which is attributed to the step edges of the (211) surface and the large lattice mismatch between SnTe and ZnSe, similar to the cases of the ZnTe/GaAs (211)[42] and ZnTe/Si (211)[43] systems. Figure S3o, p show the FFT pattern and the corresponding drawing of the atomic arrangement (211)-oriented ZnSe, and Figure S3q, r display the FFT pattern and drawing of the (211)-oriented SnTe grown on the (211)-oriented ZnSe. All the three relationships of the $[111]_{SnTe}//[001]_{Bi_2Te_3:Fe}$, $[001]_{SnTe}//[001]_{GaAs}$, $[211]_{SnTe}//[211]_{ZnSe}$ mentioned above are consistent with the HRXRD profiles shown in Figure 1.

Atomic force microscopy (AFM) images of the surfaces and the profile analyses of the three SnTe samples are shown in **Figure 2**. Figure 2a-c show the AFM images of the surface of the SnTe (111) sample with increasing magnification. Figure 2d shows the profile analyses along the green and blue lines in Figure 2c for 5 different positions. Triangular domains and terraces are clearly seen on this surface as shown in Figure 2a-c. As shown in the profile analyses in Figure 2d, for the layers forming the terraces, their heights are found to be either around 1 nm or 0.37 nm, which corresponds to the thickness of a single quintuple layer of $Bi_2Te_3$ or the step height of one SnTe (111) bilayer (BL) (0.365 nm).[44] The terraces with the step height of 1nm provide the evidence that the SnTe (111) layer follows the morphology of the triangular domains of the $Bi_2Te_3$:Fe (001) surface, while the terraces with step height around 0.37 nm indicate that the growth of SnTe (111) also occurs in a BL-by-BL manner. Figure 2e-g show the AFM images of the surface of the SnTe (001) sample with increasing magnification.

Figure 2h shows the profile analyses along the green and blue lines. It can be seen that the surface of SnTe (001) exhibits a valley structure with holes and ditches. Such a surface morphology is believed to be resulted from an initial island growth mode followed by the coalescence of some preferred islands into continuous stripes, similar to the observations by Ishikawa *et al*. [45] The profile analyses carried out for some typical holes and ditches are shown in Figure 2h. It was found that the diameters of the holes and the widths of the ditches have sizes ranging from 120 to 210 nm, and the depths of the holes and the ditches ranges from 5 to 25 nm. Figure 2i-k show the AFM images of the surface of the SnTe (211) sample with increasing magnification, which exhibits that the surface morphology of this sample is dominated by coalescing islands. Figure 2l shows the profile analysis along the green line in Figure 2k, indicating that the surface roughness can reach as high as 50 nm.

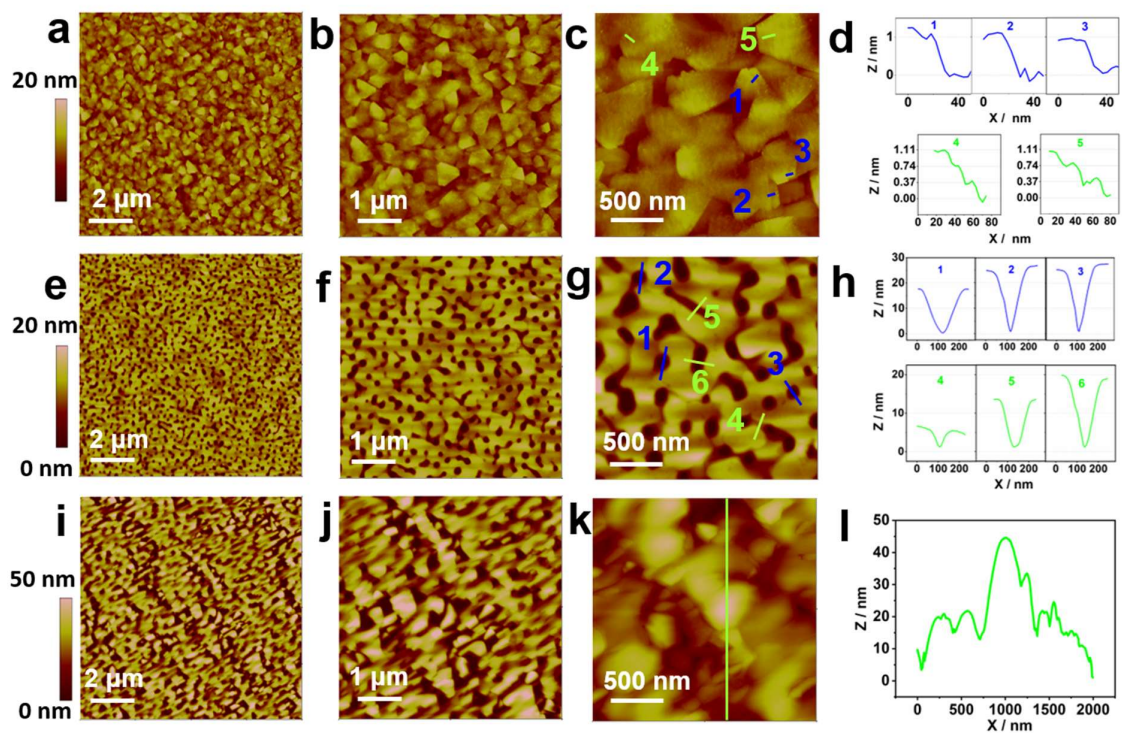

**Figure 2.** Morphology analysis of the three SnTe samples. Atomic force microscopy (AFM) images of the MBE-grown (a-c) SnTe (111), (e-g) SnTe (001), and (i-k) SnTe (211) samples. The profile analyses of (c, g, k) along the green and blue lines are presented in (d, h, l) respectively.

**Electrocatalytic HER Performance of the SnTe (001), (111) and (211) Samples**

We have examined the electrochemical catalytic HER behaviors of the as-grown SnTe (001), (111), (211) samples and a commercial Pt foil for comparison, and the results are presented in **Figure 3**. Figure 3a displays the linear sweep voltammogram (LSV) polarization curves of these materials, and as shown in Table 1, the overpotentials ($\eta$) of the SnTe (001), SnTe (111) and SnTe (211) samples at a cathodic current density ($j$) of 10 mA cm$^{-2}$ are 198 mV, 307 mV, 366 mV, respectively. The SnTe (001) thin film exhibits the lowest $\eta$ among them, even lower than the value of 48 nm Bi$_2$Te$_3$ thin film reported previously by our group,[16] the latter is among the best values of HER performances when compared with reported TI materials (exfoliated Bi$_2$Te$_3$, Bi$_2$Se$_3$, Sb$_2$Te$_3$, Bi$_{0.5}$Sb$_{1.5}$Te$_3$ alloy), and some molybdenum dichalcogenide (MoX$_2$) based nanosheet catalysts.

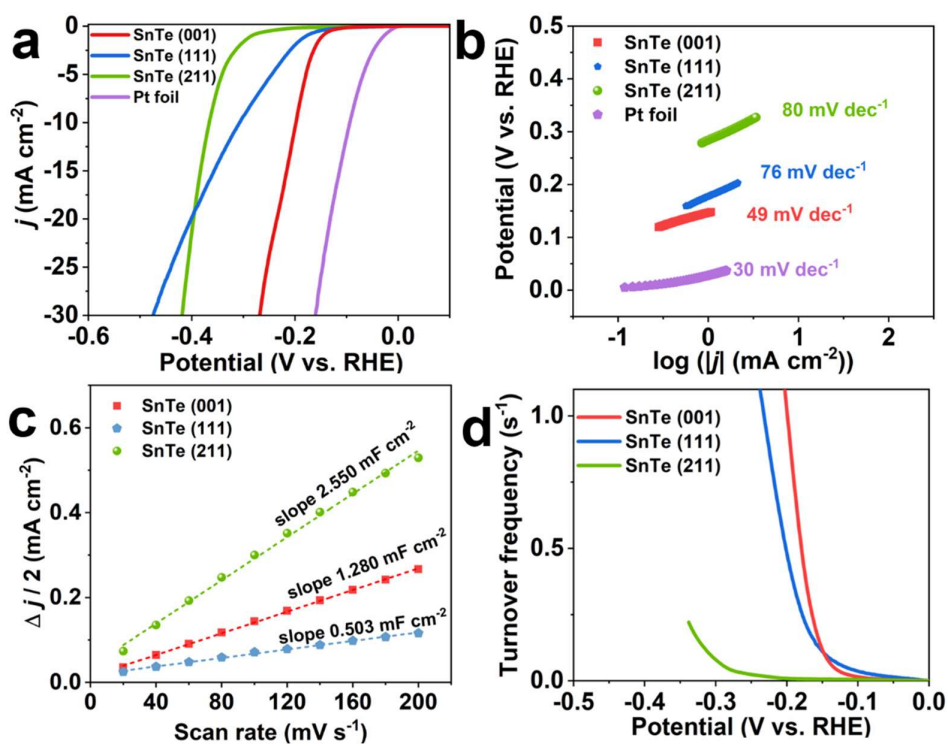

**Figure 3.** HER electrocatalytic performances of the SnTe (001), SnTe (111) and SnTe (211) samples. (a) Polarization curves (*iR*-corrected) of the three SnTe samples and a commercial Pt foil. (b) Corresponding Tafel plots of the materials in (a). (c) Linear fits of the half capacitive currents as a function of scan rates for the extraction of *C*$_{dl}$ values of the three SnTe samples. (d) Turnover frequency (TOF) plots against the potential for the three SnTe samples.

**Table 1.** Comparison of the HER performance of the three SnTe (111), SnTe (001) and SnTe (211) samples.

| Samples | Overpotential (mV vs. RHE) | Tafel slope (mV/ dec) | Exchange current density ($j_0$) ( μA cm$^{-2}$) | Log $j_0$ (A cm$^{-2}$) | $R_{ct}$ (Ω) | TOF at 200 mV (s$^{-1}$) |
|---|---|---|---|---|---|---|
| SnTe (001) | 198 | 49 | 1.065 | -5.973 | 19.84 | 1.034 |
| SnTe (111) | 307 | 76 | 4.592 | -5.338 | 23.38 | 0.482 |
| SnTe (211) | 366 | 80 | 0.288 | -6.541 | 44.50 | 0.009 |

The kinetics of active catalytic materials during the HER process can be revealed by the Tafel slope.[46] For HER in acidic media, two different mechanisms, Volmer–Tafel and Volmer–Heyrovsky, have been recognized to be responsible for transforming $H^+$ to $H_2$, which include three principal steps:

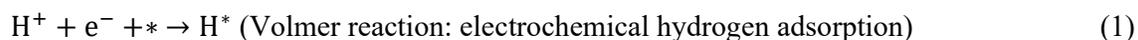

$H^+ + e^- + * \rightarrow H^*$ (Volmer reaction: electrochemical hydrogen adsorption)     (1)

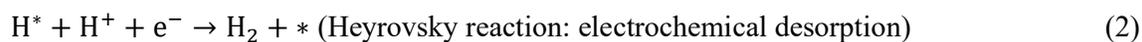

$H^* + H^+ + e^- \rightarrow H_2 + *$ (Heyrovsky reaction: electrochemical desorption)     (2)

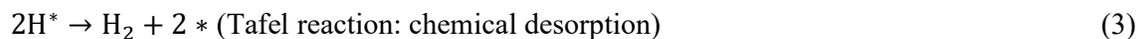

$2H^* \rightarrow H_2 + 2 *$ (Tafel reaction: chemical desorption)     (3)

where * denotes a site on the electrode surface.[47] The hydrogen evolution pathway and the rate-determining step can be deduced from the value of the Tafel slope, which is about 120, 40, or 30 mV/dec if the Volmer, Heyrovsky, or Tafel step is the rate-determining step, respectively.[46] The corresponding Tafel plots of the active materials shown in Figure 3a are displayed in Figure 3b. The linear portions of the plots shown in Figure 3b are fitted to the

Tafel equation ($\eta = b \log j + a$, where $j$ is the current density and $b$ is the Tafel slope; the

fitting equations are displayed in Figure S5). As shown in Figure 3b, the Tafel slope of the Pt

foil as a reference is 30 mV/dec which is consistent with the expected value of the known

HER mechanism on Pt, in which the HER occurs through a Volmer-Tafel mechanism and the

overall reaction is determined by the Tafel step. As shown in Figure 3b and Table 1, for the

SnTe (001), SnTe (111) and SnTe (211) samples, the respective Tafel slopes of 49 mV/dec,

76 mV/dec and 80 mV/dec suggest a Volmer-Heyrovsky mechanism, despite of different rate-

determining steps. For the SnTe (111) and SnTe (211) samples, the overall HER rate is

determined by a mixed Volmer-Heyrovsky step due to the Tafel slopes relatively close to the

average value of that of these two steps.[48-50] The fact that the SnTe (001) sample has a lower

Tafel slope of 49 mV/dec implies that its HER process is determined mainly by a typical

Heyrovsky step rather than a coupled Volmer-Heyrovsky step.[50,51] This value is close to that

of $Bi_2Te_3$ (48 nm) (47.87 mV/dec) we reported previously, which has already been among the

best value of the other TI materials and some $MoX_2$–based catalysts. The Tafel slope of the

SnTe (001) sample is the lowest among the three SnTe samples, which is consistent with the

LSV results shown in Figure 3a. Furthermore, the exchange current densities ($j_0$) were

obtained from extrapolating the x-intercepts ($\eta = 0$) of the fitted dash lines shown in Figure

S5. The obtained values of $j_0$ and the corresponding log $j_0$ (A/cm$^2$) which is an important

parameter for constructing the volcano plot [47] are shown in Table 1. The log $j_0$ (A/cm$^2$) of the

SnTe (001) sample (–5.973) and the SnTe (111) sample (–5.338) are in the same order, while the log $j_0$ (A/cm$^2$) of the SnTe (211) sample (–6.541) is significantly lower.

To gain a better understanding of the interface reactions and electrode kinetics mechanism,

electrochemical impedance spectroscopy (EIS) measurements were also performed. The

Nyquist plots of the SnTe (001), SnTe (111) and SnTe (211) samples are given in Figure S6.

We can derive the charge transfer resistance ($R_{ct}$) of them by applying a simplified Randles circuit model as shown in the inset of Figure S6. The resulted $R_{ct}$ values are shown in Table 1 in which one can see that the SnTe (001) sample exhibits the lowest $R_{ct}$ value of 19.84 $\Omega$, while that of the SnTe (111) sample and the SnTe (211) sample are 23.38 $\Omega$ and 44.50 $\Omega$, respectively, which is consistent with the observation that among the three SnTe samples, the values of $\eta$ and Tafel slope of the SnTe (211) sample are the highest and $j_0$ is the lowest. We believe the higher performance in HER (representing by the values of $\eta$ and Tafel slope) of the SnTe (001) sample over that of the SnTe (111) sample may be partially attributed to the valley surface structures with holes and ditches of the former sample as shown in the AFM images in Figure 2, which offers more active areas. However, the SnTe (211) sample exhibits the roughest surface morphology (as shown in Figure 2) but the worst electrocatalytic activity among the three SnTe samples. Thus, further HER characteristic analysis about their intrinsic electrocatalytic activity is thus essential, which has been studied and the results will be addressed below.

It is well known that the intrinsic electrocatalytic activity of a catalytic surface can be evaluated by extracting the TOF, which requires one to determine the ratio between the number of total hydrogen turnover per $cm^2$ and the number of active sites per $cm^2$.[52] A detailed description of this method and our calculations are provided in Supplementary note 1. Figure 3c shows the plots of $\Delta j/2$ versus scan rate of the three SnTe samples, in which the slopes representing the double-layer capacitance ($C_{dl}$) were extracted to be 2.550 mF $cm^{-2}$, 1.280 mF $cm^{-2}$ and 0.503 mF $cm^{-2}$ for the SnTe (211), (001) and (111) samples, respectively. The electrochemical active surface areas (ECSA) were estimated from the $C_{dl}$ values as shown in Supplementary note 1. The estimated ECSA values are consistent with the AFM image analysis as shown in Figure 2 regarding the degree of surface roughness. Figure 3d shows the TOF values versus the potentials for the three SnTe samples. Apparently, the TOF values of the SnTe (001) and SnTe (111) samples are comparable but much larger than that of the SnTe (211) sample.

As an example, the long-term stability of the SnTe (001) sample as an HER electrocatalyst was evaluated via the cyclic stability test conducted at a scan rate of 100 mV s$^{-1}$. Figure S8 shows the polarization curves recorded before and after 1000 cycles of cyclic voltammetry (CV) treatments. As can be seen in Figure S8, there is an observed potential increase of only 18.9 mV at a cathodic current density $j = 10$ mA cm$^{-2}$ after the CV treatments, representing that the electrochemical HER process does not cause significant loss in the electrocatalytic performance of the SnTe (001) sample. This slight degradation may arise from that some active components may be exfoliated during the CV treatments due to the disturbance from the evolution of H$_2$ bubbles generated on the SnTe (001) thin-film cathode. [53]

**First-Principles Calculations**

To understand the mechanism of the high catalytic efficiency induced by metallic TSSs on SnTe surfaces, we applied the DFT method to calculate the free energy of hydrogen adsorption ($\Delta G_{\mathrm{H}}$) (see Methods), [47,54,55] which is a good descriptor for the performance of HER catalysts. Either strong binding ($\Delta G_{\mathrm{H}} \ll 0\ eV$) or strong repulsion ($\Delta G_{\mathrm{H}} \gg 0\ eV$) hurts the catalytic performance. The optimal value for $\Delta G_{\mathrm{H}}$ is close to 0 eV. For example, $\Delta G_{\mathrm{H}}$ of Pt is -0.18 eV as calculated by the Perdew–Burke–Ernzerhof (PBE) exchange-correlation functional when the hydrogen coverage on the (111) surface is 1/4 monolayer (ML). [56,57]

**Figure 4** shows the $\Delta G_{\mathrm{H}}$ values for pure, Sn-vacancy-containing, and partially oxidized surfaces of SnTe. We considered the effects of hydrogen coverage, Sn vacancy, and oxygen concentrations by varying the size of unit cells and the number of oxygen atoms. In the following, we mainly compare the $\Delta G_{\mathrm{H}}$ values when the hydrogen coverage is 1/4 ML. On either (001) or Te-terminated (111) surface, $\Delta G_{\mathrm{H}}$ is 0.60 eV, which indicates that the pure

SnTe surfaces strongly repel the H atom and they do not work well for HER (we discussed the Sn-terminated (111) surface in Supplementary note 2). In fact, previous studies suggested that the experimental SnTe films were not perfect, but might contain many Sn vacancies [27,58] and were also partially oxidized.[59,60] After introducing the Sn vacancy, $\Delta G_H$ decreases considerably to 0.04 eV for the (001) surface and 0.01 eV for the (111) surface, both of which are closer to zero than $\Delta G_H$ for the (211) surface with the Sn vacancy (0.12 eV). Partial oxidation of SnTe surfaces may also provide extra active sites for H adsorption. As shown in Figure 4, $\Delta G_H$ for the partially oxidized (001) surface is reduced to -0.11 eV, much closer to zero than $\Delta G_H$ for the partially oxidized (211) surface (0.15 eV). Partial oxidation on the (111) surface also reduces $\Delta G_H$ to 0.27 eV. If the partial oxidation of a SnTe surface dominates over Sn vacancies and the surface roughness factor is not taken into account, it is thus expected that the SnTe (001) surface may offer a higher HER performance than the SnTe (111) surface. Overall, our DFT calculations show that the Sn vacancy and partial oxidation considerably decrease $\Delta G_H$, and thus enhance the electrocatalytic performance of SnTe. The decrease of $\Delta G_H$ is more apparent for the (001) and (111) surfaces than for the (211) surface.

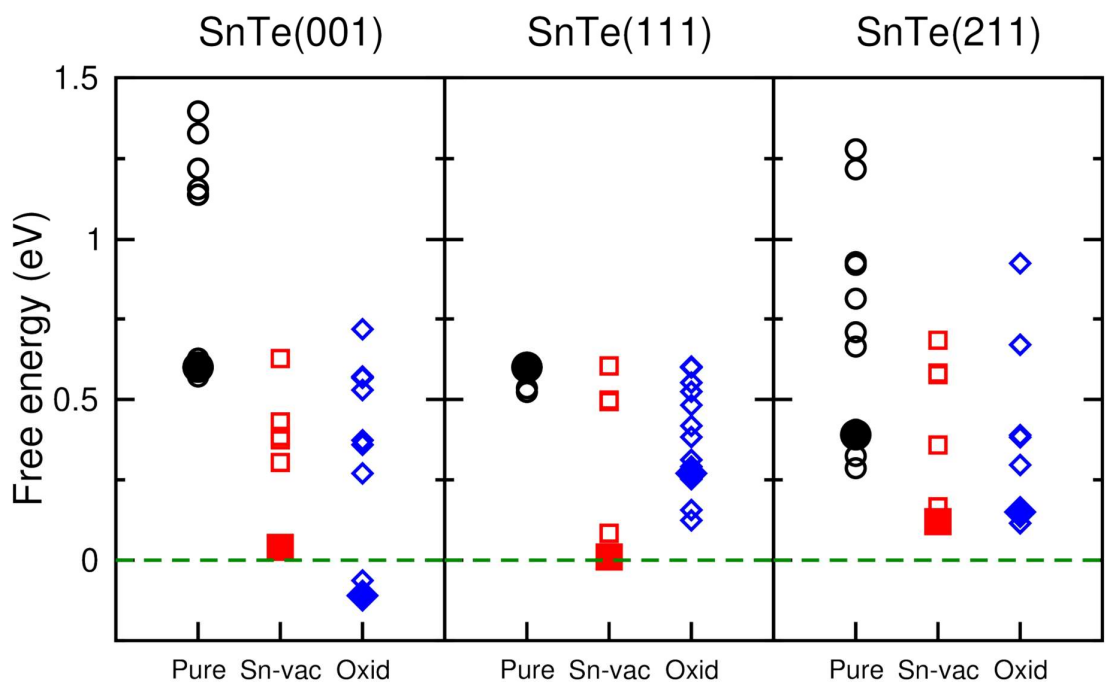

**Figure 4.** Free energies of hydrogen adsorption ($\Delta G_H$) on the SnTe (001), (111), and (211) slabs. There are three types of surfaces: pure (black circles), Sn-vacancy-containing (red squares), and partially oxidized (blue diamonds). The hydrogen coverage is between 1/9 ML and 1 ML. The filled symbols show the lowest ($\Delta G_H$) with hydrogen coverage of 1/4 ML. The atomic structures are shown in Figure S9-S17.

We performed electronic structure calculations to understand the enhancement of the electrocatalytic performances of SnTe. **Figure 5** shows that the Dirac cone shifts to a higher energy level after introducing an Sn vacancy site on the (001) surface compared to the pure surface, which makes SnTe a p-type material.[27,61] Consequently, the projected density of states (PDOS) on either Sn or Te near the Fermi level largely increases (Figure S18), which prompts the chemical bonds between the H atom and the Sn-vacancy-containing surfaces, and thus $\Delta G_H$ decreases considerably as compared with the pure surface adsorption (Table S2). When the H atom is adsorbed on the pure surfaces, the Dirac cone upshifts due to the electron transfer from TSSs to the adsorbed H atom (see Figure 5a and S19a), which does not favor a stable chemical bond. As a comparison, the H adsorption on the Sn-vacancy-containing surfaces makes the Dirac cone downshift slightly (see Figure 5b and S19b), consistent with the bond formation between the H atom and the surfaces. The Löwdin population analysis [62] (Table S3) also shows that the charge is transferred from the H atom to the (001) surface, so the H atom becomes easier to attach the Sn-vacancy surface as compared with the pure case. For the (111) surface, the Dirac cone has a similar upshift after introducing the Sn vacancy,[63] leading to a significant increase of PDOS on Te atom near the Fermi level, which also makes the H adsorption easier (Figure S20a, b). As a comparison, the (211) surface without TSSs does not have the Dirac cone, so the decrease of $\Delta G_H$ is mainly due to dangling bond states near the Fermi level, which we will discuss later.

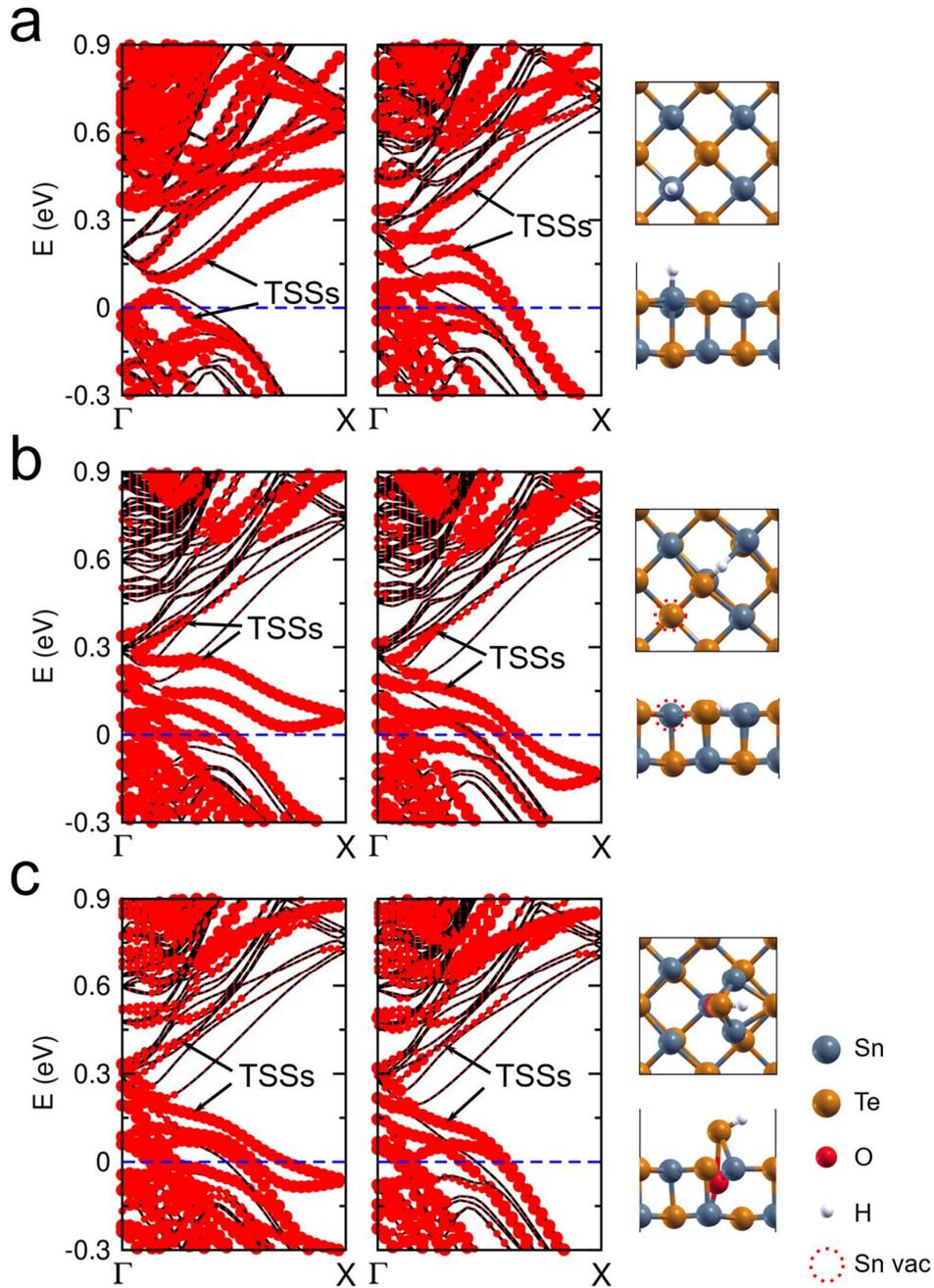

**Figure 5.** Band structures of various SnTe basal slabs before (left panel) and after (middle panel) hydrogen adsorption, and the top and side views of the surface slabs (right panel). (a) pure, (b) Sn-vacancy-containing, and (c) partially oxidized surfaces of the SnTe (001) 2 × 2 slab. The sizes of red dots represent the contributions from the (001) upper surface. The Fermi levels of the slabs are set to zero.

Our electronic structure calculations show that the surface oxidation plays a similar role as the Sn vacancy in the H adsorption. Figure 5c shows the Dirac cone moves to a higher energy level on the partially oxidized (001) surface and downshifts slightly after the H adsorption. As shown in Figure S19c, the PDOS on Te atom increases due to the upshift of the Dirac cone (Figure S18a), which favors the electron transfer from the H atom to the (001) surface (Table S3) and leads to the formation of a H-Te bonding state below the Fermi level. Thus, $\Delta G_H$ on the partially oxidized (001) surface decreases to -0.11 eV, as shown in Figure 4. A similar enhancement of the HER efficiency was also expected on the partially oxidized (111) surface (see Figure S20c), where $\Delta G_H$ decreases from 0.60 to 0.27 eV as shown in Figure 4. In our previous work, we also found that the partial oxidation on the TI $Bi_2Te_3$ helps to improve the HER activity.[16] As a comparison, we also studied the partial oxidation on the (211) surface without TSSs, and found that $\Delta G_H$ decreases from 0.39 eV to 0.15 eV, which is due to dangling bond states as mentioned earlier.

Vacancies or surface oxidation may provide dangling bond states, which may influence the electrocatalytic activity.[16,54,64-71] To compare the effects of TSSs and dangling bond states on the catalytic activities of SnTe films, we calculated the electronic structure of the pure Te-terminated SnTe (111) slab. In Figure S21a, c, the electronic states from unsaturated Te atoms at the (111) upper surface appear at the Fermi level, and after the H adsorption they disappear and the Dirac cones recover at the $\Gamma$ and M points (Figure S21b, c). Te-terminated SnTe (111) slab still has a large $\Delta G_H$ value (0.60 eV), indicating that the dangling bond states cannot enhance the HER performance much. After introducing the Sn vacancy underneath the Te-terminated (111) surface, the Dirac cone moves up and promotes the electron transfer between the H atom and the surface, leading to the formation of the H-Te bond, which does not happen to the pure (111) surface (Figure S20a, b) (Note: On the pure (111) surface, there is no stable chemical bond, because the binding energy is positive. The peak in PDOS may come from some orbital overlaps, but does not mean any bond). Additionally, we also examined the effect of dangling bond states of the (211) surface. The dangling bond states of the (211) surface appear near the Fermi level after including the Sn vacancy (Figure S22a, b), and $\Delta G_H$ decreases from 0.39 eV to 0.12 eV, indicating that the dangling bond states may promote HER activity. However, the decrease of $\Delta G_H$ value (by 0.27 eV) caused by the Sn vacancy in the (211) surface is smaller than that caused by the Sn vacancy in the (001) or (111) surfaces (0.56 or 0.59 eV), implying that metallic TSSs play a major role in the enhancement of the HER activity.

Let us summarize our DFT calculation results. We found that the pure SnTe (001) and (111) surfaces do not favor the H adsorption, indicating that they are not good HER electrocatalysts. The Sn vacancy or the partial oxidation in the surfaces lifts the Dirac cones and facilitates the charge transfer between H atoms and the TSSs during the absorption process, which boosts the HER performance. The overall performance of the (001) and (111) surfaces with TSSs is better than that of the (211) surface without TSSs, which indicates that TSSs are of great importance in the enhancement of the HER activity, consistent with our experimental results.

**XPS Studies of the SnTe (111), (001) and (211) Samples.**

The X-ray photoelectron spectroscopy (XPS) technique is a well-known tool for studying the chemical states of the surface elements of a sample. Since our theoretical studies revealed that the efficient electrocatalytic performance of the SnTe (001) sample may come from the specific dilute oxidized structures, we have performed XPS depth-profiling on the SnTe (001) sample to study the composition of the oxides near the surface. Figure S27 shows the obtained XPS spectra near the Te 3d (Figure S27a) and Sn 3d (Figure S27b) core levels of the fresh surface and those after sputtered 30 s, 60 s and 90 s for this sample. As can be seen in the top spectra in Figure S27a, which is obtained before Ar$^+$ sputtering, a peak appears on the left-hand side of the two Te core levels. These two peaks can be assigned to Te in $TeO_2$ because their binding energies are close to the reported standard values of 576.4 eV (Te $3d_{5/2}$) and 586.8 (Te $3d_{3/2}$) of this oxide.[72,73] It can also be seen that the intensities of the two major peaks in Figure S27a increase while their full width at half maximums (FWHM) get narrower with the sputter time. As shown in Figure S27b, the value of FWHM of the two Sn core level peaks of the fresh surface (1.46 eV) is larger than what is expected from a single core level peak. In fact, after 30 s of sputtering, these two peaks can be seen to be composed of at least two peaks. As sputtering further proceeds, these two peaks become higher and narrower. These observations also indicate that the two Sn core level peaks also involve surface oxide components.

In the top spectrum of **Figure 6**a, the raw data of the two Te $3d_{5/2}$ peaks with background subtracted are displayed together with three fitted peaks. The peak positioned at 576.1 eV can be assigned to $TeO_2$, and the peak at 572.2 eV corresponds to the $Te^{2-}$ state of SnTe, which agree with their previously reported standard values.[73] It is worthwhile to note that the peak at 572.8 eV in the top graph in Figure 6a corresponds to the elemental $Te^0$ state.[73] Its existence can be explained by the mechanisms proposed by Neudachina *et al* [73] that $Te^0$ may be an intermediate for $Te^{4+}$ ($TeO_2$) formation: $Te^{2-} \xrightarrow{v_1} Te^0 \xrightarrow{v_2} Te^{4+}$. $Te^0$ accumulated at the surface is attributed to that the rate of the first step likely exceeds that of the second step ($v_1 > v_2$). It has been reported that the Te peaks of $Te^{4+}$ in $TeO_2$, elemental $Te^0$, $Te^{2-}$ in SnTe have approximately symmetric shapes.[74-78] In the top graph of Figure 6a, the raw data of the Te $3d_{5/2}$ peaks with background subtracted are shown together with three symmetric fitted curves of $Te^{4+}$, $Te^0$ and $Te^{2-}$; however, the overall fitted curve in red does not fit the raw data well. Obviously, an additional peak in the left shoulder of the $Te^0$ peak must be added to obtain the best fit of the raw data as shown in the bottom graph of Figure 6a. In the top graph of Figure 6b, the raw data of the Sn $3d_{5/2}$ peak with background subtracted are displayed together with two fitted peaks. The peak positioned at 486.7 eV can be assigned to the $Sn^{4+}$ state of $SnO_2$, and the peak at 485.3 eV corresponds to the $Sn^{2+}$ state of SnTe, which agree with their previously reported standard values.[73] It has also been reported that the $Sn^{4+}$ and $Sn^{2+}$ peaks have symmetric shapes.[79-82] In the top graph of Figure 6b, the raw data of the Sn $3d_{5/2}$ peak with background subtracted are shown together with two symmetric fitted curves of $Sn^{4+}$ and $Sn^{2+}$; however, the overall fitted curve in red does not fit the raw data well. Similarly, an additional peak is also needed to obtain a better fit as shown in the bottom graph of Figure 6b. Based on this analysis, we have performed data fitting for all the original XPS spectra by adding a peak at the left shoulder of each of the elemental $Te^0$ and $Sn^{2+}$ peaks in SnTe. Indeed, all the raw data including those obtained after sputtering are now well fitted as shown in Figure 6c,d with all the binding energies of the major peaks of $Te^{4+}$ in $TeO_2$, elemental $Te^0$, $Te^{2-}$ in SnTe, $Sn^{4+}$ in $SnO_2$, $Sn^{2+}$ in SnTe consistent with the reported standard values.[60,73,78,83,84] Interestingly, previous XPS studies [60,85] reported that oxidized Te surface and SnTe surface each contains a small suboxide peak with binding energies similar to that of the additional peaks described in Figure 6. It is believed that the two additional peaks in our XPS spectra provide the evidence of the existence of the partially oxidized SnTe surface structures predicated from our theoretical studies.

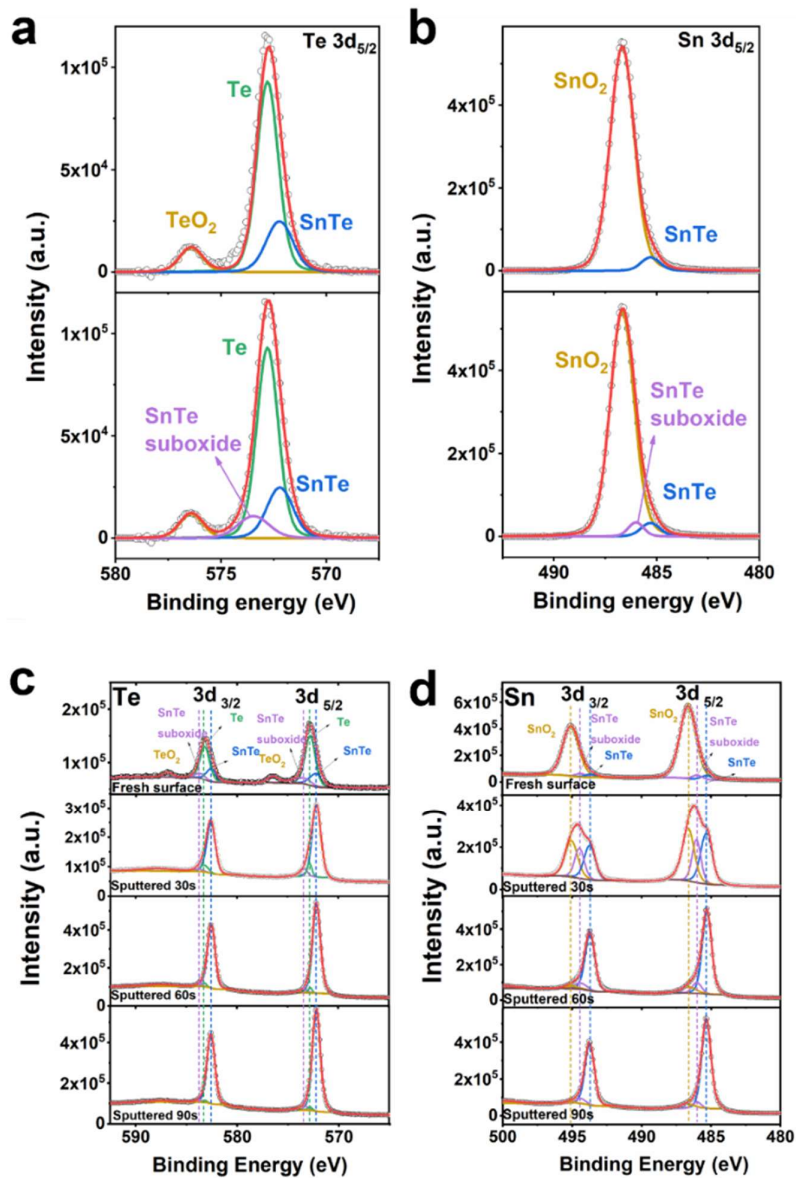

**Figure 6**. X-ray photoelectron spectroscopy (XPS) spectra obtained from the surface of the SnTe (001) sample exposed to air after the MBE growth. (a) Te $3d_{5/2}$ and (b) Sn $3d_{5/2}$ with Shirley background subtraction; the top and bottom graphs display the fitting curves before and after adding an additional peak of the SnTe suboxide structure, respectively. (c) Te 3d and (d) Sn 3d, the four graphs from top to bottom show the spectra of the fresh surface, and those after sputtered for 30 s, 60 s and 90 s, respectively, together with the fitted curves by including the peak of the SnTe suboxide structures.

As shown in the top second graph of Figure 6c, after sputtering for 30 s, the $TeO_2$ peak becomes undetectable, while the peaks associated with the elemental $Te^0$ decrease obviously, which are consistent with the mechanism mentioned above that elemental $Te^0$ is an intermediate during the formation of $TeO_2$. However, it should be noted that both the SnTe suboxide and SnTe peak increase. The bottom two graphs in Figure 6c display that further sputtering results in a gradual decrease in the intensities of the SnTe suboxide and elemental $Te^0$ peaks. As shown in the top second graph of Figure 6d, after sputtering for 30 s, the peak intensity of $SnO_2$ decreases significantly and that of both the SnTe suboxide and SnTe increase obviously. The bottom two graphs in Figure 6d show that both the $SnO_2$ and SnTe suboxide peaks drop gradually upon further sputtering. It is important to mention that the SnTe suboxide peaks of Te and Sn core levels show higher intensities after the surface was sputtered for the first 30 s as mentioned above as compared to that of the fresh surface, which likely indicates that these SnTe suboxide structures are mainly located below the major oxides of $Te^0$, $TeO_2$ and $SnO_2$. It is worthwhile to mention that the electrolyte used in our electrochemical measurements is a 0.5 M $H_2SO_4$ solution, which can etch away these major oxides, such that the SnTe surfaces with Sn vacancies and partially oxidized structures are exposed to make the H adsorption easier and enhance the HER activity.

**WAL effect on the SnTe (001), (111) and (211) samples.**

The TSSs in TCI SnTe can be experimentally probed by low-temperature magneto-transport measurements. WAL effect due to spin-momentum locking and accumulation of the π Berry phase in magneto-transport is considered as a characteristic feature of TSSs. In general, the transport of TSSs, as nontrivial two-dimensional (2D) states, usually occur concurrently with that of the trivial 3D bulk states.[86] However, TSSs lead to the 2D WAL effect in perpendicular applied magnetic field while 3D bulk states give rise to the 3D WAL effect, which is independent of the direction of the magnetic field.[87] The WAL effect due to TSSs can be obtained by subtracting the WAL effect in parallel applied magnetic field from that in perpendicular applied magnetic field. As shown in **Figure 7**a–c, the SnTe samples with the surface orientation of (001), (111) and (211) all exhibit sharp dips in the magnetoresistance $(MR(\%) = \frac{R(B) - R(0)}{R(0)} \times 100\%)$ around zero magnetic field when magnetic field is applied in both perpendicular and parallel directions. The sharp dip is a hallmark of the WAL effect due to quantum correction to the classical quadratic magnetoresistance. Figure 7d–f show that after subtracting the 3D bulk contribution, the WAL effect due to TSSs in terms of magnetoconductance $\Delta G = \Delta G(B_\perp) - \Delta G(B_{//})$ only remains in the SnTe (001) and SnTe (111) samples. No noticeable TSSs-induced WAL feature for the SnTe (211) sample is observed.[86] These results are consistent with the fact that TSSs protected by crystal symmetry exist and are robust against disorder in the highly symmetric surfaces of SnTe.[25,26] The high HER performance of the SnTe (001) and SnTe (111) samples observed in this work is thus believed to be attributed to the TSSs. Absence or fragility of TSSs in the lowly symmetric SnTe (211) seemingly leads to the observed inferior catalytic activity.

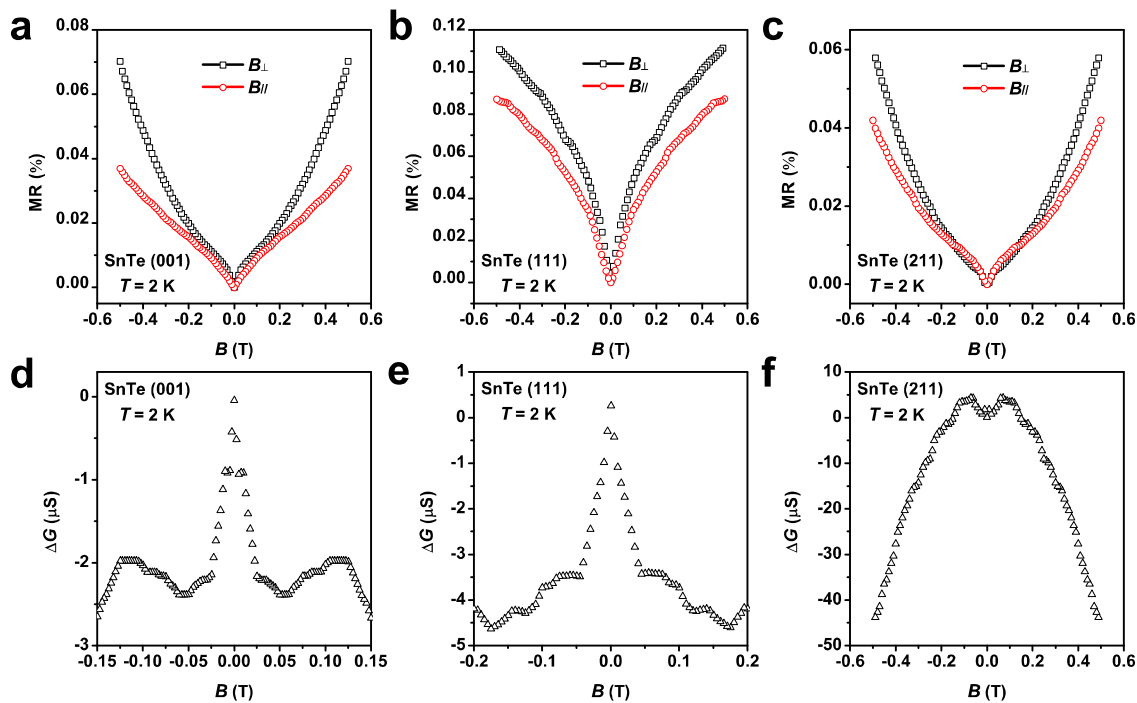

**Figure 7**. WAL effect observed in the SnTe (001), SnTe (111) and SnTe (211) samples at 2 K. (a–c) Magnetoresistance in (a) SnTe (001), (b) SnTe (111) and (c) SnTe (211) under perpendicular ($B\perp$) and parallel ($B_{//}$) magnetic field. (d–f) Magnetoconductance calculated by the subtraction $\Delta G(B\perp) - \Delta G(B_{//})$ as a function of magnetic field of (d) SnTe (001), (e) SnTe (111) and (f) SnTe (211).

**Further discussion on the role of the TSSs**

In the previous sections, we have shown that even the SnTe (211) sample has the roughest surface, however, its HER performance is the lowest among the three SnTe samples studied in our work. Someone may argue that this observation may be simply due to the generation of various facets at these rough surfaces and they may have different intrinsic activities. Certainly, a rough surface could bring out facets with various orientations at the upper most surface of a sample. As shown in our recent work[16], among five Bi$_2$Te$_3$(001) thin films with different thicknesses, it is likely that they also contain surface facets with various orientations, and the 48 nm thin film with the roughest surface enjoys the highest HER performance. This observation indeed could rule out the facet factor being a significant negative contribution while supporting that a rougher surface could have more active sites for HER. For the three SnTe samples studied in this work, as described earlier regarding the TOF values, the dominating factor among the three SnTe samples in their HER performance is not the number of available active sites, instead it is about how active of these sites among the three samples, while their intrinsic activity is dominated by the TSSs contribution. In fact it has a big contrast between the (211) oriented thin film and the other two thin films as shown in our theoretical calculations regarding their $\Delta G_H$ values. In addition, the TOF values as shown in Figure 3d and Table 1 indicate that the intrinsic activities of the (001) and (111) samples could be 50 to 100 times higher than that of the (211) sample at potential of 200 mV. It is well known that TSSs can float to the top of the sample surface and they are robust and can survive against the surface defects or non-full oxidation and as demonstrated by other researchers theoretically[23] and by us experimentally[16,22]. For the TCI SnTe, the TSSs come from the crystalline interface underlying the rough surface as protected by the mirror symmetry. Thus even though the (211) surface has the highest roughness corresponding to the highest number of active sites, however, these sites are much less active, therefore the positive contribution from the higher number of active sites is overwhelmed by the TSSs effect that is only available and robust for the (001) and (111) orientations among the three orientations studied.

**Conclusion**

We have successfully synthesized high-quality SnTe (111), (001), and (211) samples on GaAs substrates by the MBE technique. Their structural properties were examined by RHEED, HRXRD, cross-sectional HRTEM, and AFM. Their performance in HER was characterized by various electrochemical measurements. It was found that the SnTe (001) sample exhibits the lowest $\eta$, the lowest Tafel slope, and the lowest $R_{et}$ among the three SnTe samples. As revealed by AFM analysis, the three surfaces show very different surface morphologies. Using the LSV and CV measurements, their intrinsic electrochemical activities were extracted with the results that SnTe (001) and (111) surfaces outperform the (211) surface. We have carried out DFT calculations to study the underlying mechanism of our experimental observations. We found that while the pure SnTe surfaces do not favor the H adsorption, the Sn vacancy or the partial oxidation in the surfaces lifts the Dirac cones and makes the H adsorption easier, which boosts the HER performance. The overall performance of the (001) and (111) surfaces with TSSs is better than that of the (211) surface without TSSs. These theoretical findings were further supported by XPS analysis and magneto-transport measurements. The issues about the effect of possible surface facets and the contrast in the activity of their active sites among the three samples were also addressed, which further confirm the important role of the TSSs effect that is only available and robust for the (001) and (111) oriented samples. This work demonstrates that the TSSs and mirror symmetry of the SnTe (001) and (111) surfaces play an important role in their efficient HER performance and shines the light in achieving cost-effective electrocatalysts in water splitting based on the use of TCIs.

**Methods**

**Sample preparation**

All the SnTe (001), (111) and (211) samples studied in this work were fabricated on n+ GaAs substrates by a VG-V80H MBE system equipped with *in situ* RHEED. Sample synthesis was conducted using a high-purity SnTe compound source. For the SnTe (001) sample, a SnTe layer was deposited directly on the GaAs (001) substrate at substrate temperature of 222 °C. For the SnTe (111) sample, a 5 nm ZnSe buffer layer was first deposited, followed by the growth of a $Bi_2Te_3$:Fe layer at substrate temperature of 242 °C (a lower temperature of 234 °C was used for the first 5 min, it was found that this two-step growth mode provided a better structural quality) and a SnTe layer at substrate temperature of 222 °C. For the SnTe (211) sample, a 630 nm ZnSe buffer layer was deposited firstly, and then a 120 nm SnTe layer was deposited at substrate temperature of 222 °C. All the growth processes were performed in an ultrahigh-vacuum chamber with a basic pressure better than $1.0 \times 10^{-9}$ Torr.

**Material characterization**

All the SnTe (001), (111) and (211) thin film samples were characterized by HRXRD (PANalytical multipurpose diffractometer using Cu K$\alpha_1$ X-rays with a wavelength of 1.54056 Å), TEM (JEOL JEM-2010F with an acceleration voltage of 200 kV), and AFM (Dimension 3100 with a NanoScope IIIa controller (Digital Instruments) using the tapping mode). All the XPS measurements were performed using Kratos-Axis Ultra DLD XPS *ex situ*. This instrument was equipped with a monochromatic Al K$\alpha$ X-ray source (photon energy 1486.7 eV, 150 W), and the measurements were taken in hybrid lens mode with an energy step of 100 meV, a pass energy of 40 eV, and a large measuring area of $1 \times 2$ mm$^2$. The ion sputtering of the film was handled using Ar ions with 4 kV, 3 mm $\times$ 3 mm raster, 140 µA extractor current, and the sputtering rate was found to be similar to that of SiO$_2$, $\sim$1Å s$^{-1}$.

**Electrochemical Measurements for HER**

All the electrochemical measurements were performed in a standard three-electrode electrolyzer connected to a CHI 660E electrochemical workstation (CH Instruments), using SnTe (001), (111) or (211) as the working electrode, respectively, a graphite rod as a counter electrode, a standard Ag/AgCl electrode (saturated KCl solution) served as the reference electrode, and 0.5 M H$_2$SO$_4$ as the electrolyte (degassed by N$_2$, purity $\sim$99.995%). In constructing the working electrode, a piece of the as-grown sample is connected to a poly-ether-ether-ketone(PEEK) electrode via a conductive glassy carbon clip, one side of the clip touches the back side of the GaAs substrate and the other side connects a small part of the sample surface to the PEEK electrode. As mentioned in our previous work[16], the GaAs substrate has negligible HER activity. LSV was performed using a scan rate of 5 mV s$^{-1}$. EIS measurements were carried out at open circuit potential over a frequency range of $10^6$ to 0.01 Hz with a perturbation voltage amplitude of 5 mV. The impedance data were fitted to a simplified Randles circuit to extract the series and charge-transfer resistances. All data presented were *iR* corrected. The potential values shown were with respect to the reversible hydrogen electrode.

**Magneto-transport characterizations**

Electrical contacts were prepared by thermal evaporation of Cr/Au thin films with typical thicknesses of 10 nm/100 nm through shadow masks on SnTe (001), (111) and (211) thin films. The samples were subsequently patterned into four-terminal devices. The magneto-transport measurements were performed in a Quantum Design Physical Property Measurement System (PPMS, Quantum Design) equipped with a rotational sample holder.

**Theoretical Calculations**

We conducted DFT calculations with plane-wave basis sets and the PBE exchange-correlation functional[88] implemented in the Quantum Espresso package (v.6.1).[89] We applied the Optimized Norm-Conserving Vanderbilt (ONCV) pseudopotentials[90-92] for H, Sn and Te and ultrasoft pseudopotential[93] for O, and the kinetic energy cutoff is 50 Ry. We did spin-orbit coupling (SOC) calculations with full relativistic pseudopotentials to identify the gapless TSSs. An experimental lattice constant of 6.328 Å was adopted for the SnTe rock salt structure.[58,94,95] In the structural optimization, the upper 2–monolayer (ML) with the adsorbed H atom could move until the force on each atom was smaller than 0.0001 Ry/bohr, and the other atoms were fixed at the bulk positions. In electronic structure calculations, the surface structures without H adsorption were obtained by directly removing the H atom without any structural relaxation. With periodic boundary conditions, we kept at least 12 Å vacuum to avoid interactions between neighboring replicas. We used the $6 \times 6 \times 1$ Monkhorst-Pack k-point mesh[96] in the structural optimization of the $(1 \times 1)$ SnTe (001) unit cell, and $8 \times 8 \times 1$ for electronic structure calculations with SOC. More detailed computational-model configurations are included in Supplementary note 3.

**Data availability**

The data that support the findings of this study are available from the corresponding author upon reasonable request.

ASSOCIATED CONTENT

**Supporting Information.** High resolution X-ray diffraction 2θ-ω scan of a SnTe thin film layer directly deposited on a GaAs (111) substrate. The lower part shows the powder diffraction files (PDFs) of SnTe and GaAs as references. Sample structures and reflection high-energy electron diffraction (RHEED) studies of the three SnTe samples. (a-c) Sample structure and RHEED patterns of $Bi_2Te_3$ (0001) taken when the electron beam is along the $[10\bar{1}0]$ and $[2\bar{1}\bar{1}0]$ direction and SnTe (111) taken when the electron beam is along the $[2\bar{1}\bar{1}]$ and $[10\bar{1}]$ direction of the SnTe (111) sample. (d, e) Sample structure and RHEED patterns of SnTe (001) taken when the electron beam is along the [100] and [110] direction of the SnTe (001) sample. (f, g) Sample structure and RHEED pattern of SnTe (211) taken when the electron beam is along the $[01\bar{1}]$ direction of the SnTe (211) sample. Schematic drawings at the bottom parts of (b, c, e) are the top views of the respective surface lattices, where the electron beam is oriented directly upward in the page. Cross-sectional high-resolution transmission electron microscopy (HRTEM) images and their structural analysis of the three MBE-grown (a-f) SnTe (111), (g-l) SnTe (001), and (m-r) SnTe (211) samples. Cross-sectional high-resolution TEM images of (a-b) SnTe (111), (g-h) SnTe (001) and (m-n) SnTe (211) samples. (c, e, i, k, o, q) The corresponding fast Fourier transform (FFT) patterns of $Bi_2Te_3$: Fe (001), SnTe (111), GaAs (001), SnTe (001), ZnSe (211) and SnTe (211), respectively. (d, f, j, l, p, r) Schematic drawings of atomic arrangements of $Bi_2Te_3$:Fe (001), SnTe (111), GaAs (001), SnTe (001), ZnSe (211) and SnTe (211) respectively. Energy dispersive spectroscopy (EDS) profile of the $Bi_2Te_3$:Fe layer indicates that the incorporated Fe concentration is about 0.80% (Table S1). . Exchange current densities ($j_0$) for the active materials in Figure 3a can be derived from the x-intercepts of the fitted dash lines extrapolated from the Tafel plots shown in colors.  Nyquist plots of the SnTe (001), SnTe (111) and SnTe (211) samples. Electrochemical surface area measurements. Cyclic voltammograms (CV) curves with different scan rates of the (a) SnTe (001), (b) SnTe (111) and (c) SnTe (211) samples. Polarization curves ($iR$-corrected) of the SnTe (001) sample recorded before and after 1000 cycles of cyclic voltammetry (CV) using accelerated degradation tests (scan rate = 100 mV s$^{-1}$). Structures of H adsorption on pure SnTe (001) surface; Structures of H adsorption on Sn-vacancy-containing SnTe (001) surface; Structures of H adsorption on partially oxidized SnTe (001) surface; Structures of H adsorption on pure SnTe (111) surface; Structures of H adsorption on Sn-vacancy-containing SnTe (111) surface; Structures of H adsorption on partially oxidized SnTe (111) surface; Structures of H adsorption on pure SnTe (211) surface; Structures of H adsorption on Sn-vacancy-containing SnTe (211) surface; Structures of H adsorption on partially oxidized SnTe (211) surface; Projected density of states (PDOS) on the Sn (blue) and Te (red) atoms in the (a) pure and (b) Sn-vacancy-containing SnTe (001) surfaces; PDOS on hydrogen-hosting sites on (a) pure, (b) Sn-vacancy containing, and (c) partially oxidized SnTe (001) surfaces before (blue) and after (red) hydrogen adsorption; PDOS on the hydrogen-hosting Te atom on (a) pure, (b) Sn-vacancy-containing, and (c) partially oxidized SnTe (111) surface before (blue) and after (red) hydrogen adsorption; Band structures of  (1 × 1) 48ML SnTe (111) slab (a) before and (b) after hydrogen adsorption and the PDOS on Te atom before (blue) and after (red) hydrogen adsorption with 1ML coverage; PDOS on the Te and Sn atoms in (a) pure and (b) Sn-vacancy-containing and (c) partially oxidized SnTe (211) surfaces before (blue) and after (red) hydrogen adsorption; Band structure of pure $(2 \times 1)$-1Sn (111) surface with 39 ML thickness; PDOS on hydrogen-hosting Te atom on (a) pure and (b) partially oxidized $(2 \times 1)$-1Sn (111) surfaces before (blue) and after (red) hydrogen adsorption; Structures of H adsorption on $(2 \times 1)$-1Sn (111) surface; Structures of H adsorption on partially oxidized $(2 \times 1)$-1Sn (111) surface; Raw data of Te 3d (a) and Sn 3d (b) XPS spectra of the SnTe (001) sample for its fresh surface, and those after sputtering for 30s, 60s, and 90s. Composition of the $Bi_2Te_3$:Fe layer of the SnTe (111) sample determined by EDS (**Table S1**). $\Delta G_H$ values on a Sn atom, a Te atom, and a Sn-Te bridge on pure and Sn-vacancy-containing surfaces of a $(2 \times 2)$ SnTe (001) slab (**Table S2**); Löwdin charges of the hydrogen atom adsorbed on pure, Sn-vacancy-containing, and partially oxidized surfaces of SnTe (001), (111), and (211) slabs (**Table S3**); Measurement of the electrochemical specific capacitance, Calculation of the electrochemical active surface area,  Turnover frequency calculations (Supplementary note 1); Hydrogen adsorption on Sn-terminated (111) surface (Supplementary note 2); Methods for theoretical calculations (Supplementary note 3).

AUTHOR INFORMATION

Corresponding Author

*E-mail: dingpan@ust.hk (D.P.).

*E-mail: phiksou@ust.hk (I.K.S).

Author Contributions

Q.Q. and B.L. contributed equally to this work. Q.Q. and I.K.S. initiated this study and further designed the experiments; Q.Q. carried out the sample synthesis, conducted the structural characterizations and electrochemical measurements; J.L. helped to do the analyses of RHEED patterns and TEM images. B.L. and D.P. carried out the theoretical calculations; H.T.L. and J.N.W conducted the magneto-transport measurements. Q.Q., I.K.S., D.P., B.L., H.T.L. and J.N.W wrote the manuscript. All authors performed the data analysis and discussions.

Notes

The authors declare no competing financial interest.

ACKNOWLEDGMENT

This research was funded by the Research Grants Council of the Hong Kong Special Administrative Region, China, under Grant Numbers 16304515, 16301418, C6013-16E, 16308020, C6025-19G and William Mong Institute of Nano Science and Technology under Project Number WMINST19SC06 and WMINST19SC07. D.P. acknowledges support from the Croucher Foundation through the Croucher Innovation Award and the Energy Institute at the Hong Kong University of Science and Technology.

ToC figure:

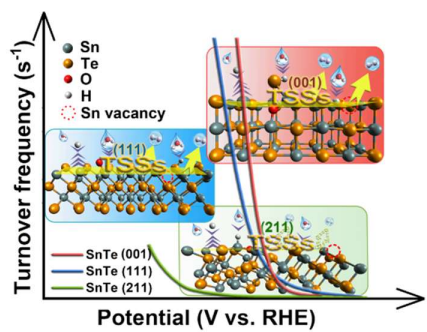

# Supporting Information

**Role of topological surface states and mirror symmetry in topological crystalline insulator SnTe as an efficient electrocatalyst**

*Qing Qu, Bin Liu, Hongtao Liu, Jing Liang, Jiannong Wang, Ding Pan\* and Iam Keong Sou\**

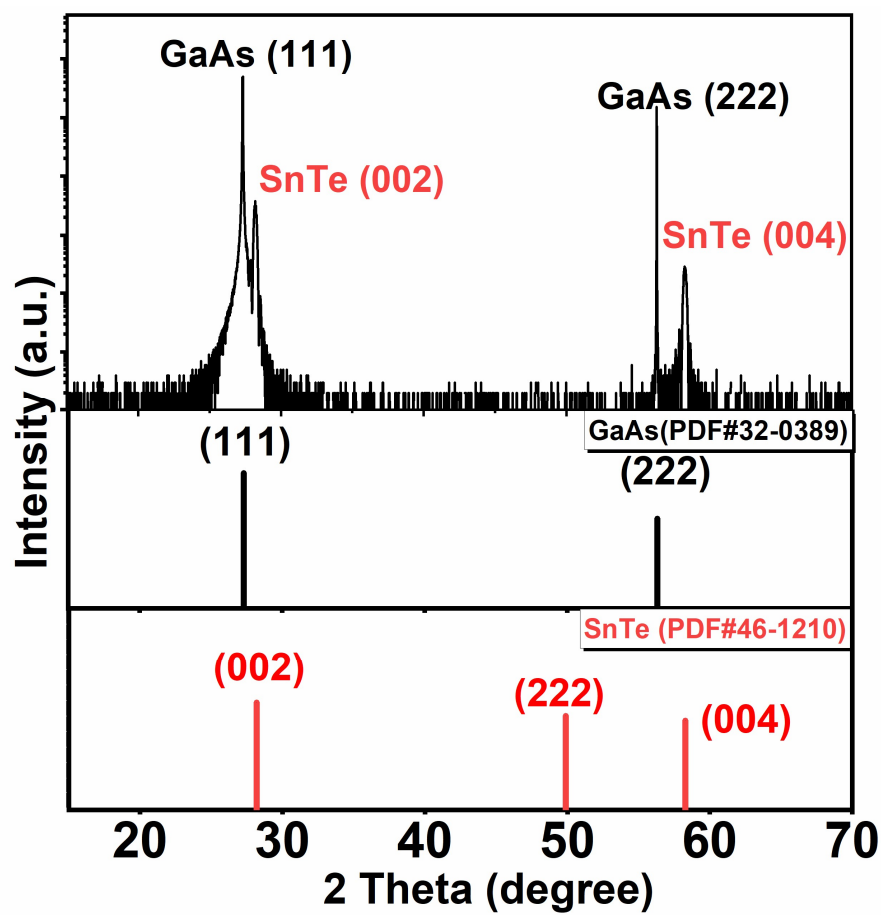

Figure S1. High resolution X-ray diffraction 2θ-ω scan of a SnTe thin film layer directly deposited on a GaAs (111) substrate. The lower part shows the powder diffraction files (PDFs) of SnTe and GaAs as references.

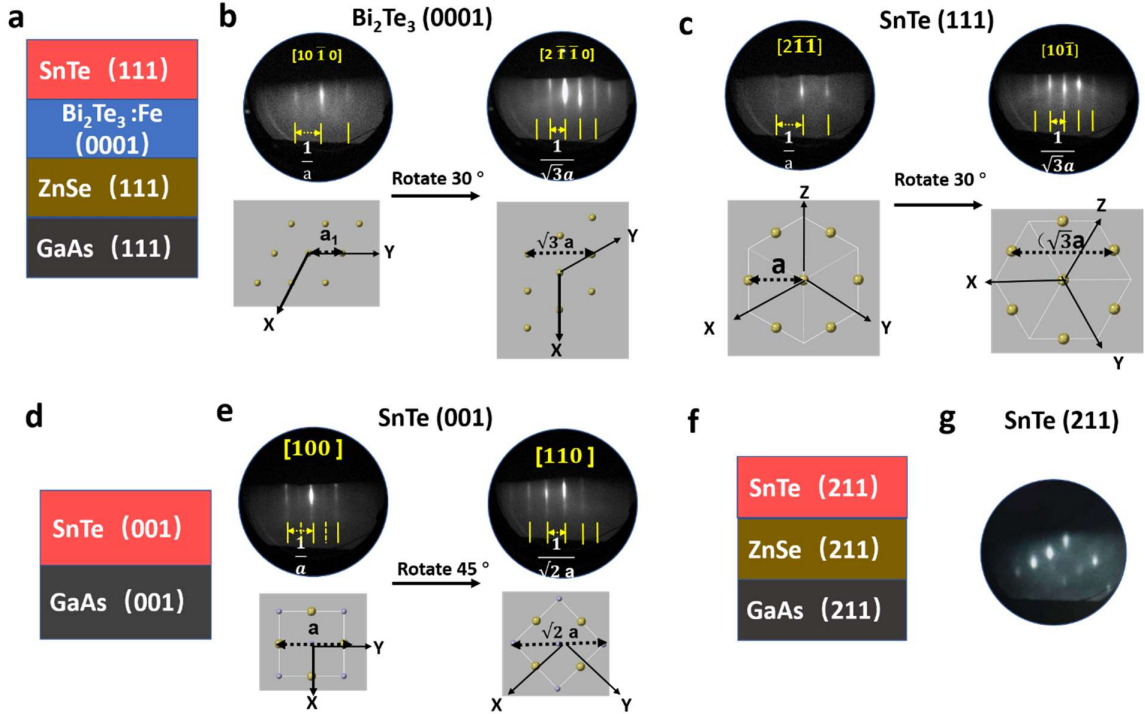

Figure S2. Sample structures and reflection high-energy electron diffraction (RHEED) studies of the three SnTe samples. (a-c) Sample structure and RHEED patterns of $Bi_2Te_3$ (0001) taken when the electron beam is along the $[10\bar{1}0]$ and $[2\bar{1}\bar{1}0]$ direction and SnTe (111) taken when the electron beam is along the $[2\bar{1}\bar{1}]$ and $[10\bar{1}]$ direction of the SnTe (111) sample. (d, e) Sample structure and RHEED patterns of SnTe (001) taken when the electron beam is along the [100] and [110] direction of the SnTe (001) sample. (f, g) Sample structure and RHEED pattern of SnTe (211) taken when the electron beam is along the $[01\bar{1}]$ direction of the SnTe (211) sample. Schematic drawings at the bottom parts of (b, c, e) are the top views of the respective surface lattices, where the electron beam is oriented directly upward in the page.

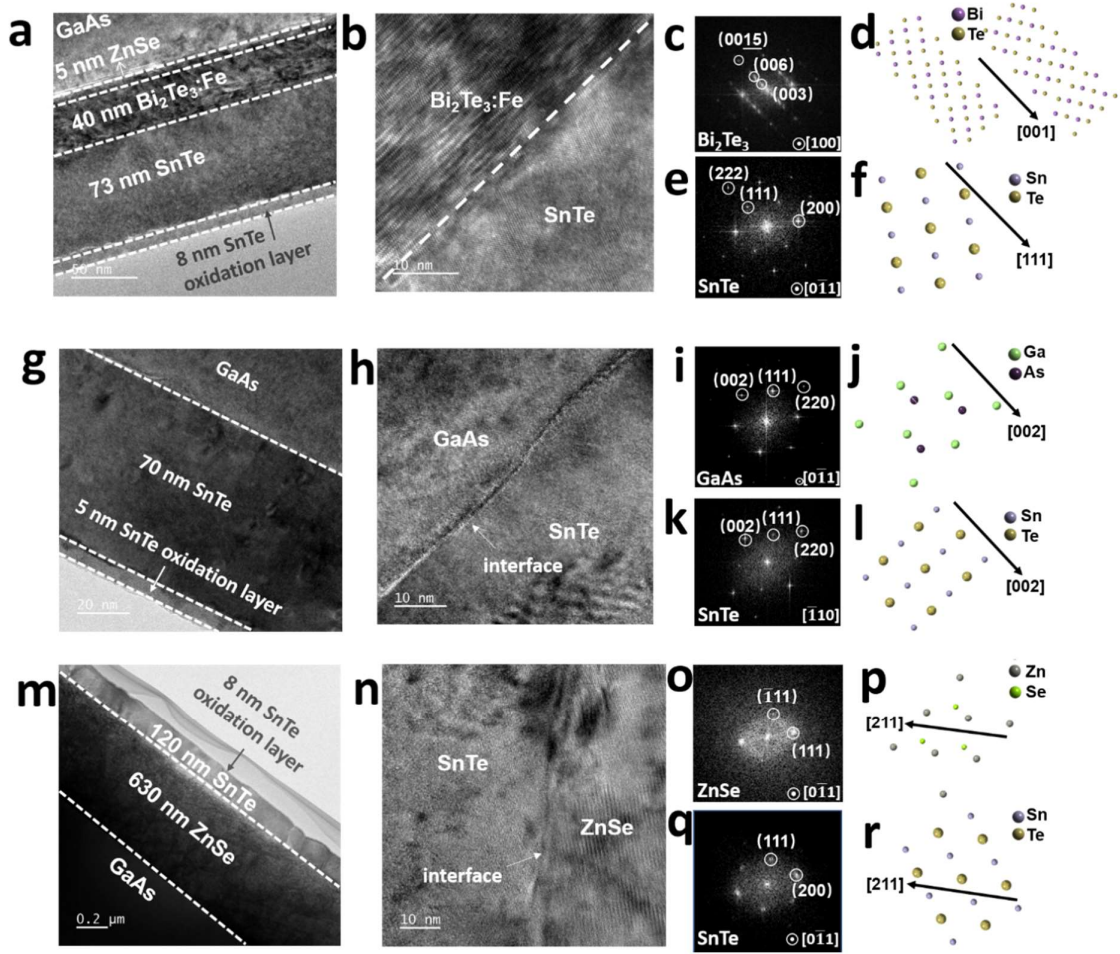

Figure S3. Cross-sectional high-resolution transmission electron microscopy (HRTEM) images and their structural analysis of the three MBE-grown (a-f) SnTe (111), (g-l) SnTe (001), and (m-r) SnTe (211) samples. Cross-sectional high-resolution TEM images of (a-b) SnTe (111), (g-h) SnTe (001) and (m-n) SnTe (211) samples. (c, e, i, k, o, q) The corresponding fast Fourier transform (FFT) patterns of $Bi_2Te_3$: Fe (001), SnTe (111), GaAs (001), SnTe (001), ZnSe (211) and SnTe (211), respectively. (d, f, j, l, p, r) Schematic drawings of atomic arrangements of $Bi_2Te_3$:Fe (001), SnTe (111), GaAs (001), SnTe (001), ZnSe (211) and SnTe (211) respectively.

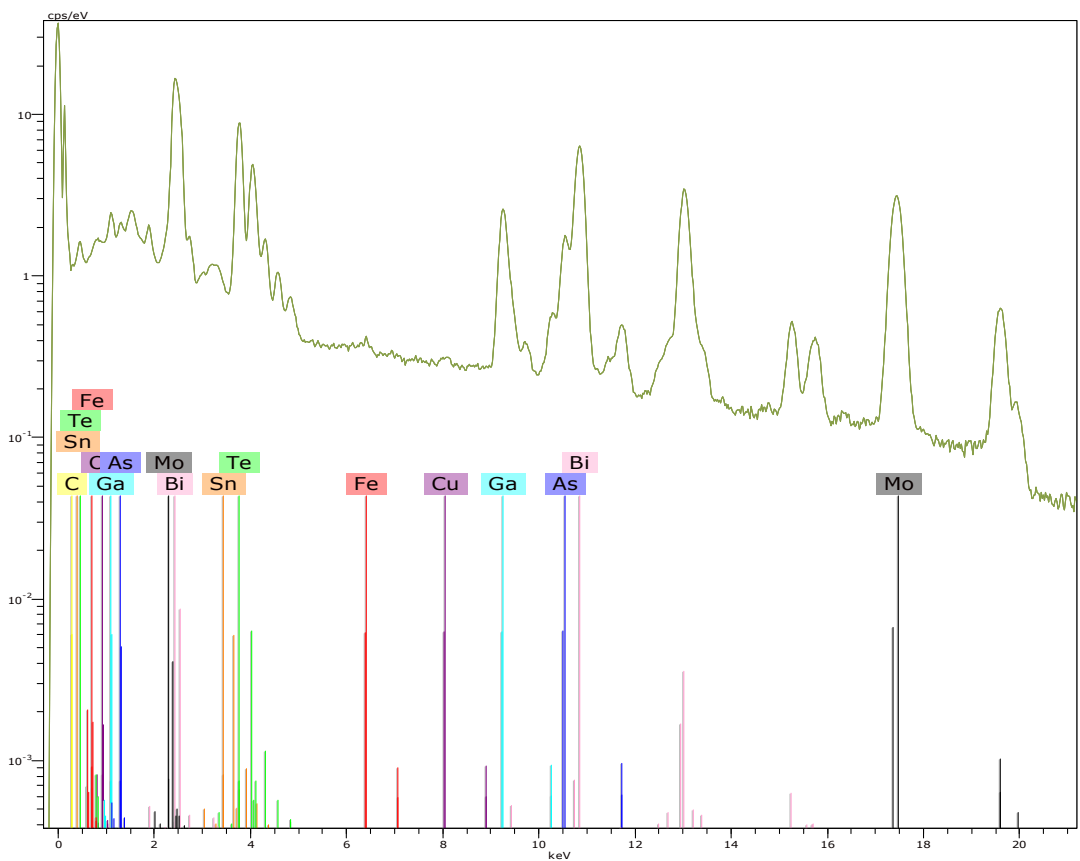

Figure S4. Energy dispersive spectroscopy (EDS) profile of the Bi$_2$Te$_3$:Fe layer indicates that the incorporated Fe concentration is about 0.80% (Table S1), where the Mo signal comes from the Mo sample holder used for the EDS measurements and the Cu signal comes from the contamination from the ion-milling chamber for TEM sample preparation as many users used Cu-grid sample holders for this process.

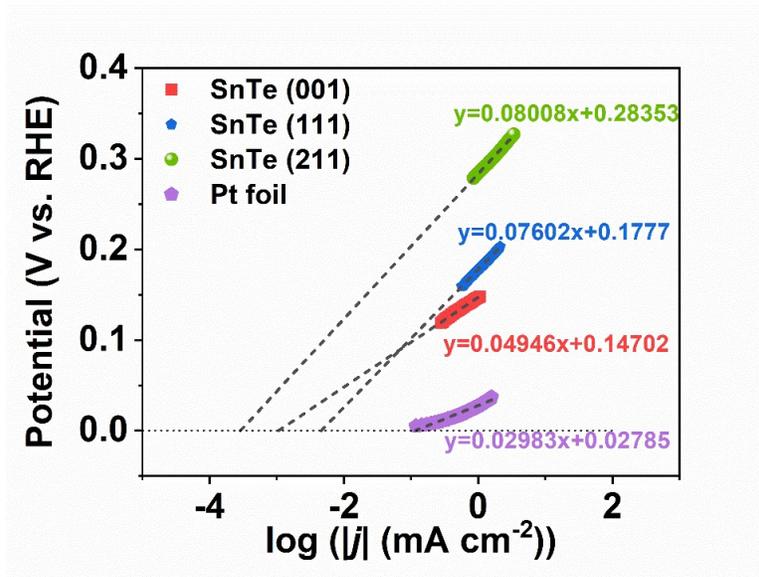

Figure S5. Exchange current densities ($j_0$) for the active materials in Figure 3a can be derived from the x-intercepts of the fitted dash lines extrapolated from the Tafel plots shown in colors.

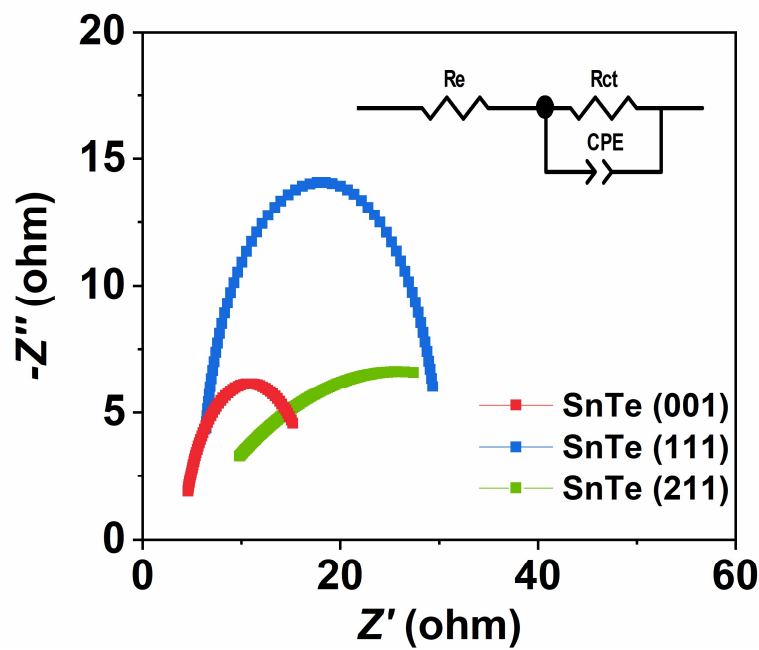

Figure S6. Nyquist plots of the SnTe (001), SnTe (111) and SnTe (211) samples.

Remarks of Figure S6:

Within the period for taking an EIS spectrum, the system being measured must be at a steady state and the potential of such a state is usually determined by finding out the open-circuit potential (OCP) of the system. OCP is the difference between the potential of the working electrode and that of the reference electrode when the cell is at open-circuit status. In our studies, the determination of the OCP value for a sample lasts for at least one hour. After inserting the sample to the electrolyte, we define the system is under static state if the OCP value changes no more than 1mV within 10 mins. For different samples, the OCP values are different as they depend on parameters like surface inhomogeneity and roughness, thus the EIS measurements are usually conducted at the sample's own OCP value.

Both the Nyquist plots of the SnTe (001) and SnTe (111) samples present typical semicircle profiles, while the Nyquist plot of SnTe (211) shows an arc of a circle with its center lying some distance below the x-axis (in the fourth quadrant). Most of the researchers describe the feature of the Nyquist plots similar to that of SnTe (211) as depressed and/or deformed semicircles [1-4]. A possible cause for such depressed semicircles is attributed to the deviation from a smooth electrode surface [5], causing an uneven current distribution on the electrode due to the surface inhomogeneous and roughness [6-10]. This observation is also consistent with the AFM image analysis shown in Figure 2 that the surface of the SnTe (211) sample exhibits the roughest surface morphology compared with the surfaces of the SnTe (001) and SnTe (111) samples.

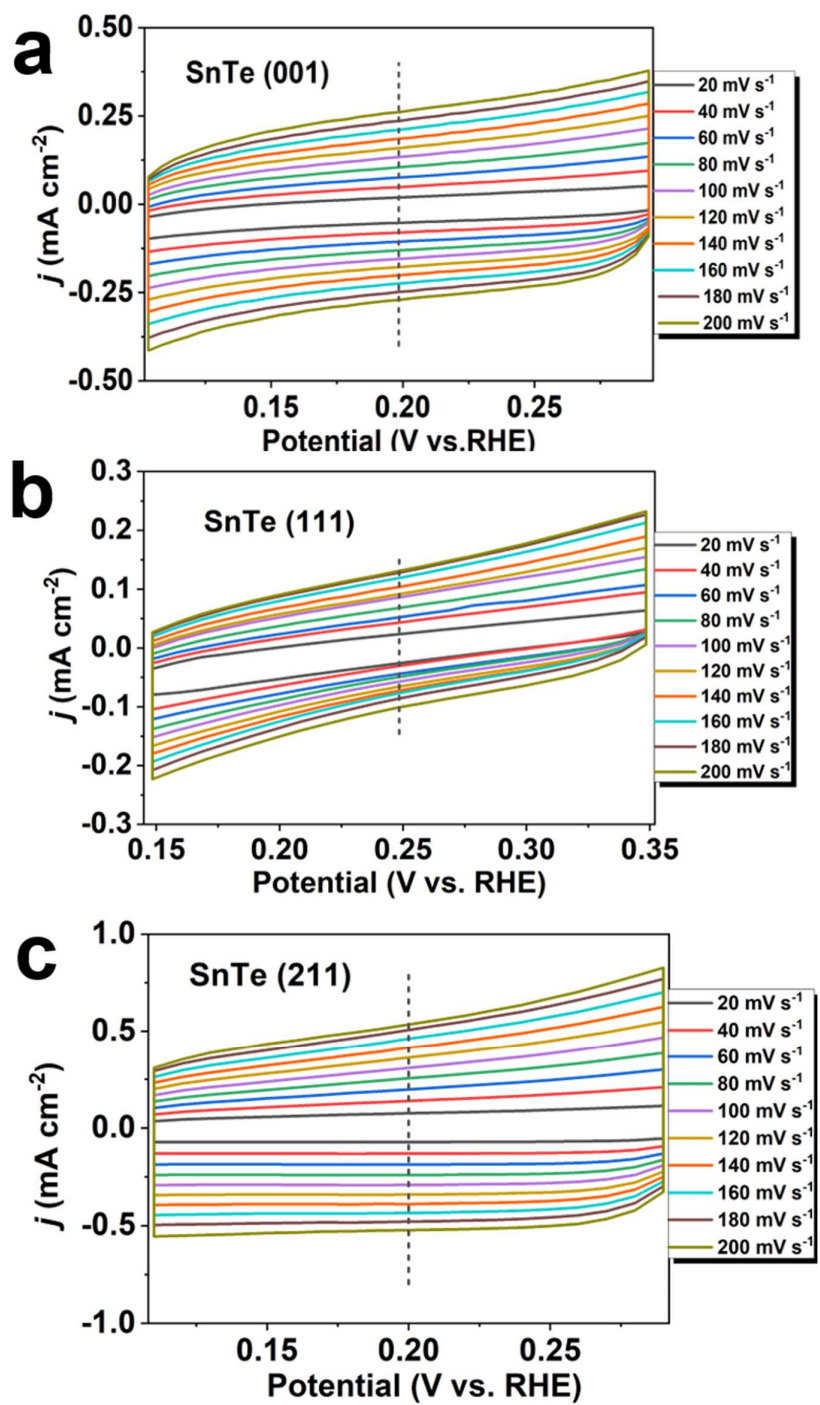

Figure S7. Electrochemical surface area measurements. Cyclic voltammograms (CV) curves with different scan rates of the (a) SnTe (001), (b) SnTe (111) and (c) SnTe (211) samples.

Supplementary note 1

**Measurement of the electrochemical specific capacitance**

To measure the electrochemical specific capacitance of the SnTe (001), (111) and (211) samples, the potentials were swept at each of ten different scan rates (20, 40, 60, 80, 100, 120, 140, 160, 180 and 200 mV/s).

Figure S7 shows the results of these cyclic voltammetry (CV) measurements. We measured the capacitive currents in the potential ranges of the curves in Figure S7 where no faradic processes are observed, and the capacitive current densities (($j_{\text{anodic}}$-$j_{\text{cathodic}}$)/2 taken at the potential values of the dash lines) are plotted as a function of the scan rate as shown in Figure 3c.

**Calculation of the electrochemical active surface area**

The $C_{\text{dl}}$ values can be converted into the electrochemically active surface areas (ECSA) using the specific capacitance value for a flat standard with 1 cm$^2$ of real surface area. The specific capacitance for a flat surface is generally found to be in the range of 20-60 µF cm$^{-2}$ [11-14]. In the following calculations for obtaining the TOF values, we take 40 µF cm$^{-2}$ as a moderate value, and define the A$_{\text{ECSA}}$ as:

$$A_{\text{ECSA}} = \frac{C_{\text{dl}}}{40 \text{ µF cm}^{-2} \text{per cm}^2_{\text{ECSA}}}$$

Thus the ECSA values of the three SnTe samples can be calculated as shown below.

$$A_{\text{ECSA}}^{\text{SnTe (001)}} = \frac{1.280 \text{ mF cm}^{-2}}{40 \text{ µF cm}^{-2} \text{per cm}^2_{\text{ECSA}}} = 32 \text{cm}^2_{\text{ECSA}}$$

$$A_{ECSA}^{SnTe\ (111)} = \frac{0.503\ \text{mF cm}^{-2}}{40\ \mu\text{F cm}^{-2}\text{per cm}_{ECSA}^2} = 12.575\ \text{cm}_{ECSA}^2$$

$$A_{ECSA}^{SnTe\ (211)} = \frac{2.55\ \text{mF cm}^{-2}}{40\ \mu\text{F cm}^{-2}\text{per cm}_{ECSA}^2} = 63.75\text{cm}_{ECSA}^2$$

**Turnover frequency calculations**

To calculate the per-site turnover frequency (TOF), we used the following formula:

$$TOF = \frac{\text{number of total hydrogen turnover/ cm}^2 \text{of geometric area}}{\text{number of active sites/ cm}^2 \text{of geometric area}}$$

The number of total hydrogen turnover can be calculated according to [15]:

$$\left(|j|\frac{\text{mA}}{\text{cm}^2}\right)\left(\frac{1A}{1000\text{mA}}\right)\left(\frac{\frac{1C}{s}}{1A}\right)\left(\frac{6.241\times10^{18}e^-}{1C}\right)\left(\frac{1H_2}{2e^-}\right) = \left(3.12\times10^{15}\frac{\frac{H_2}{s}}{\text{cm}^2}\right)|j|,$$ where $j$ is taken

from the polarization curves.

In general, the HER activity of a sample is the average activity from the various sites of the sample. For samples with rough surface, the number of the active sites per unit surface is usually estimated to be the 2/3 power of the ratio of the number of Sn and Te atoms in a SnTe unit cell over the unit volume [16-18].

The number of the active sites of the SnTe (001), (111) and (211) samples per real surfaces area can be calculated using the parameters presented below:

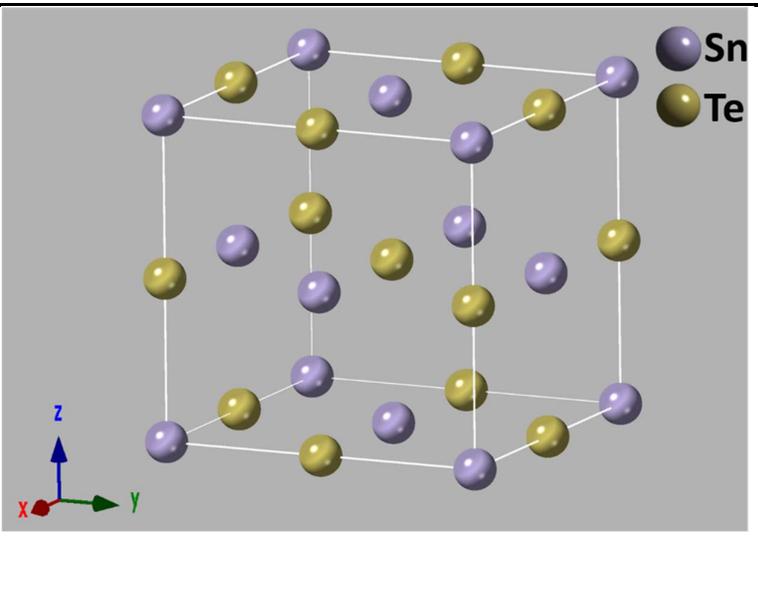

| | |
|---|---|
| SnTe unit cell | |
| Volume | 251.598 Å³ |
| Contains | 4 Te atoms, 4 Sn atoms |

$$\text{Active sites}_{\text{SnTe}} = \left(\frac{8 \text{ atoms/unit cell}}{251.598 \text{ Å}^3/\text{unit cell}}\right)^{\frac{2}{3}} = 1.004 \times 10^{15} \text{ atom cm}_{\text{real}}^{-2}$$

The TOF values of the SnTe (001), (111), and (211) samples can be calculated by the following equation:

$$\text{TOF} = \frac{3.12 \times 10^{15} \frac{\text{H}_2/\text{s}}{\text{cm}^2}}{1.004 \times 10^{15} \text{ atom cm}_{\text{real}}^{-2} \times \text{A}_{\text{ECSA}}^{\text{SnTe}}} |j|.$$

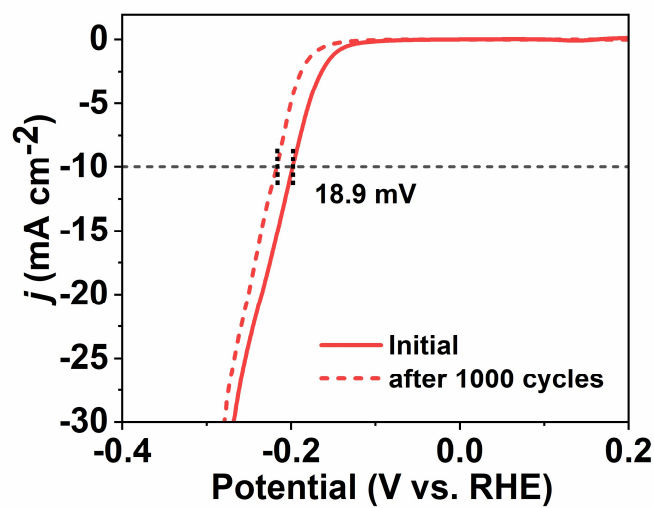

Figure S8. Polarization curves (*iR*-corrected) of the SnTe (001) sample recorded before and after 1000 cycles of cyclic voltammetry (CV) using accelerated degradation tests (scan rate = 100 mV s$^{-1}$).

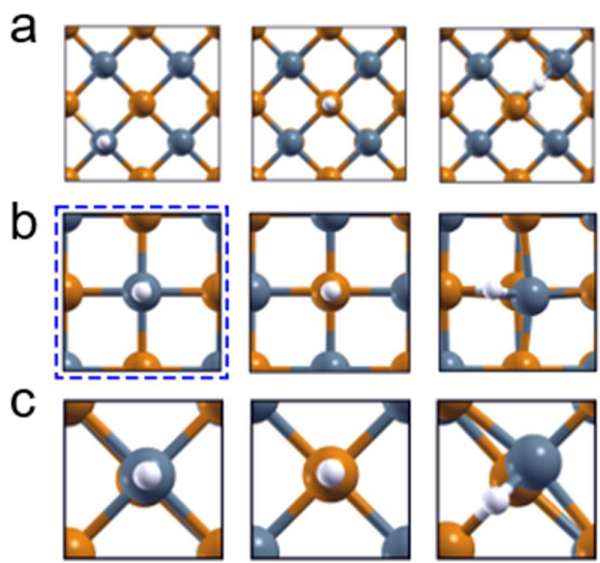

Figure S9. Top views of hydrogen adsorption on pure SnTe (001) surface at Sn, Te and Sn-Te bridge sites with hydrogen coverage of (a) 1/8 ML (b) 1/4 ML and (c) 1/2 ML. The blue dashed box highlights the most stable adsorption structure with hydrogen coverage of 1/4 ML.

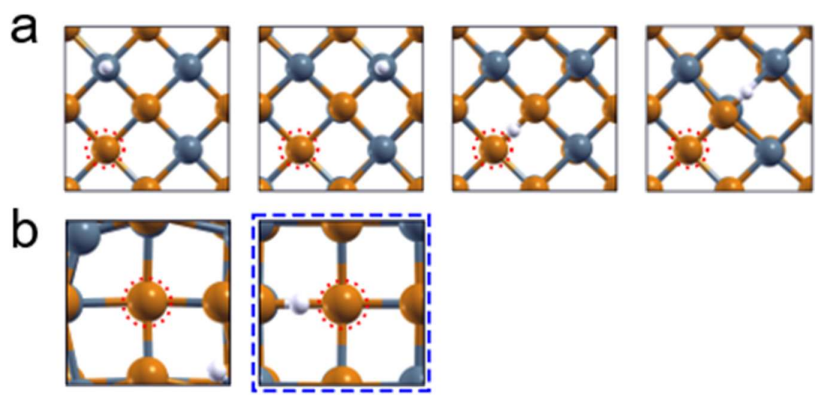

Figure S10. Top views of hydrogen adsorption on Sn-vacancy-containing SnTe (001) surface at different adsorption sites with hydrogen coverage of (a) 1/8 ML and (b) 1/4 ML. Red dashed circles indicate Sn-vacancy positions. The blue dashed box highlights the most stable adsorption structure with hydrogen coverage of 1/4 ML.

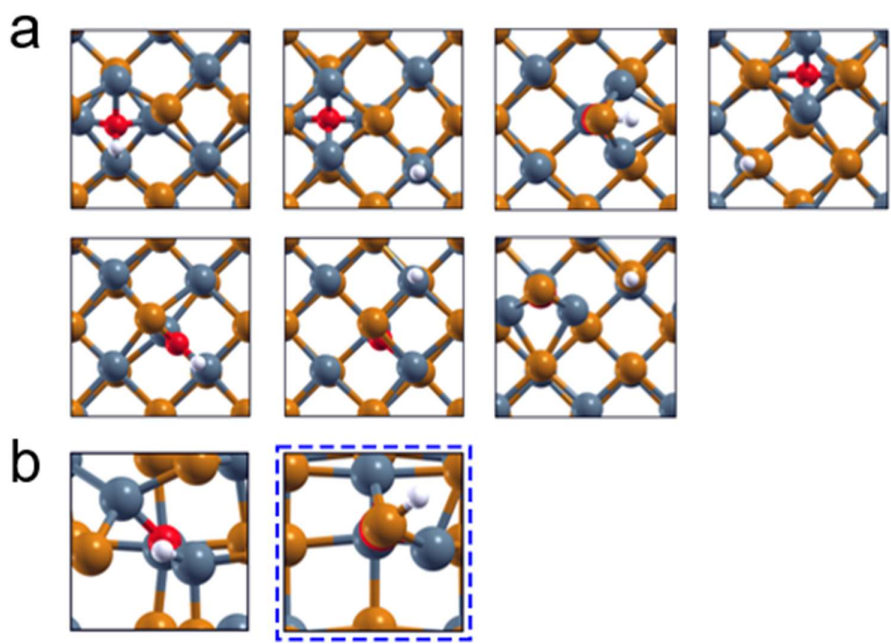

Figure S11. Top views of hydrogen adsorption on partially oxidized SnTe (001) surface at different adsorption positions with hydrogen coverage of (a) 1/8 ML and (b) 1/4 ML. The blue dashed box highlights the most stable adsorption structure with hydrogen coverage of 1/4 ML.

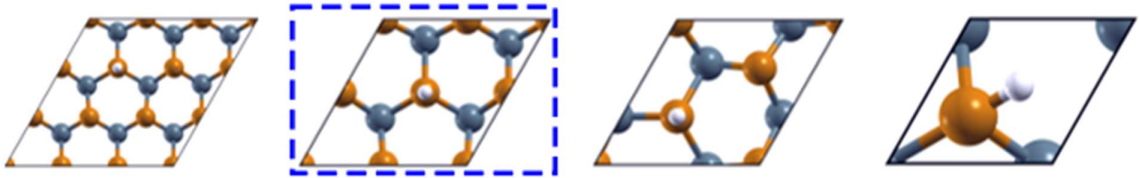

Figure S12. Top views of hydrogen adsorption on pure SnTe (111) surface with hydrogen coverage of 1/9 ML, 1/4 ML, 1/3 ML and 1 ML (from left to right). According to the surface symmetry, only one energetically stable hydrogen adsorption configuration can be obtained on each unit cell. The blue dashed box highlights the most stable adsorption structure with hydrogen coverage of 1/4 ML.

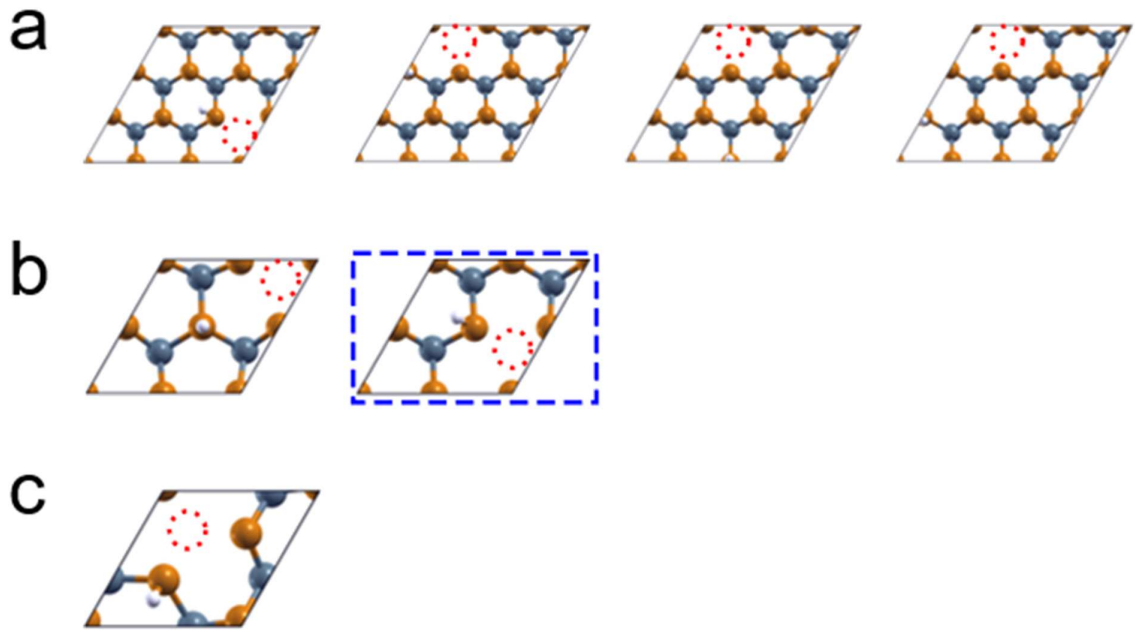

Figure S13. Top views of hydrogen adsorption on Sn-vacancy-containing SnTe (111) surface with hydrogen coverage of (a) 1/9 ML, (b) 1/4 ML and (c) 1/3 ML. Red dashed circles indicate Sn-vacancy positions. The blue dashed box highlights the most stable adsorption structure with hydrogen coverage of 1/4 ML.

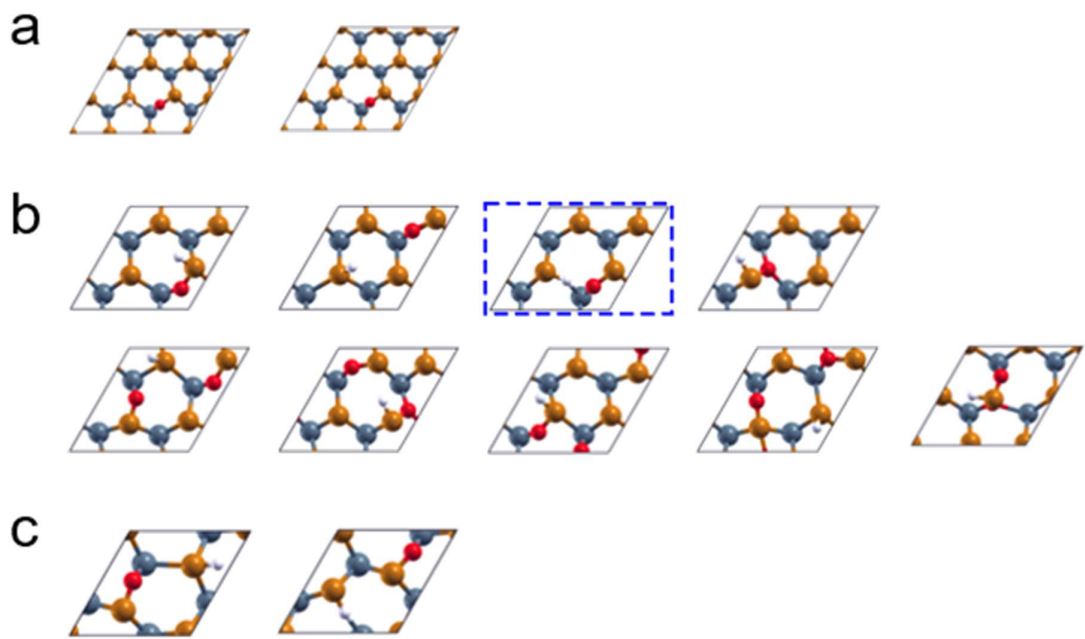

Figure S14. Top views of hydrogen adsorption on partially oxidized SnTe (111) surface at different adsorption positions with hydrogen coverage of (a) 1/9 ML, (b) 1/4 ML and (c) 1/3 ML. The upper panel of (b) includes one oxygen atom and the lower panel includes two oxygen atoms onto (111) surface. The blue dashed box highlights the most stable adsorption structure with hydrogen coverage of 1/4 ML.

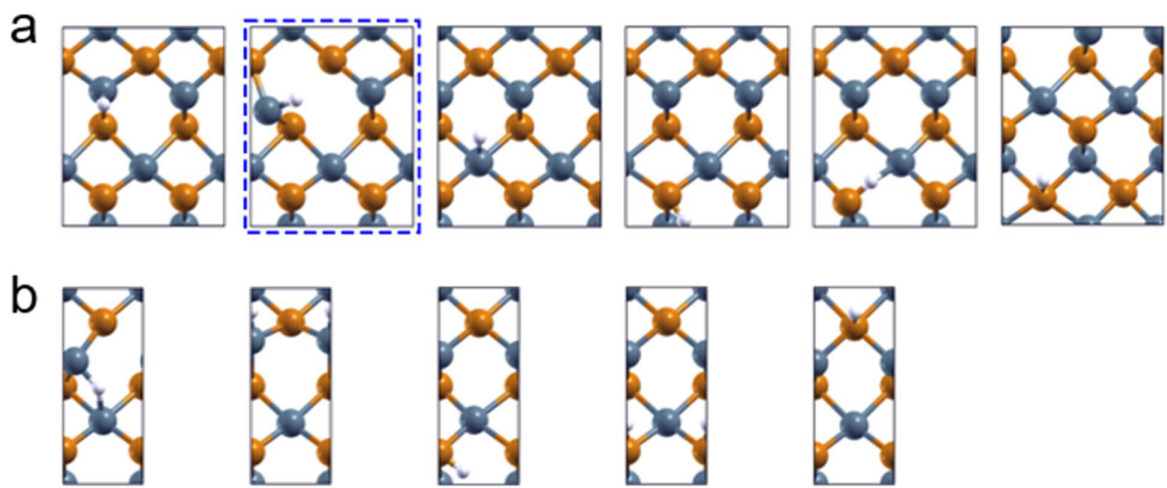

Figure S15. Top views of hydrogen adsorption on pure SnTe (211) surface at different adsorption positions with hydrogen coverage of (a) 1/4 ML and (b) 1/2 ML. The blue dashed box highlights the most stable adsorption structure with hydrogen coverage of 1/4 ML.

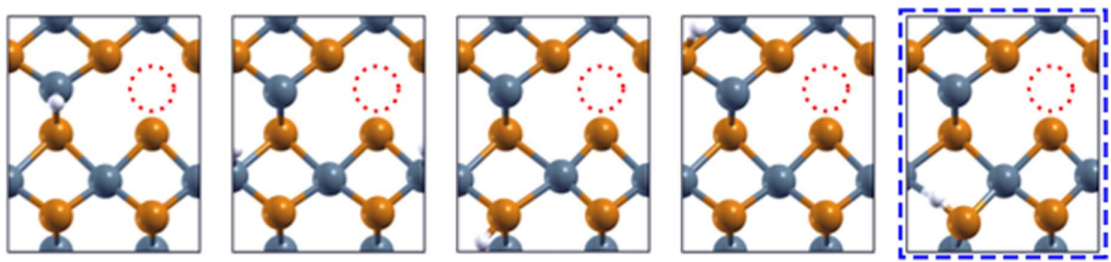

Figure S16. Top views of hydrogen adsorption on Sn-vacancy-containing SnTe (211) surface at different adsorption positions with hydrogen coverage of 1/4 ML. Red dashed circles indicate the Sn-vacancy positions. The blue dashed box highlights the most stable adsorption structure.

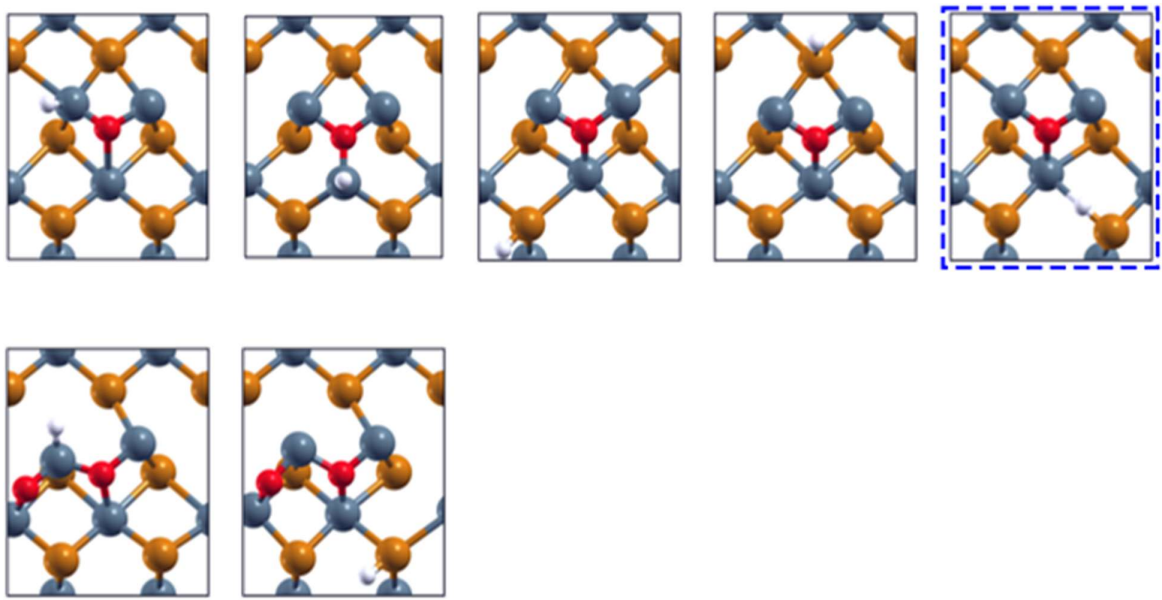

Figure S17. Top views of hydrogen adsorption on partially oxidized SnTe (211) surface at different adsorption positions with hydrogen coverage of 1/4 ML. The upper panel includes one oxygen atom and the lower panel includes two oxygen atoms onto (211) surface. The blue dashed box highlights the most stable adsorption structure.

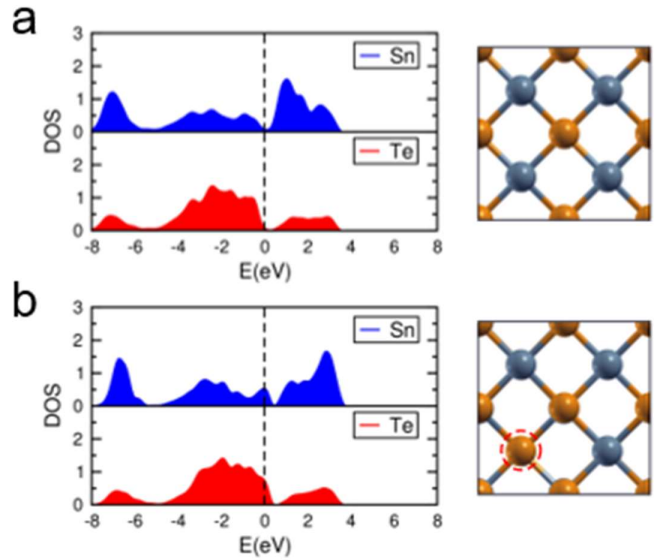

Figure S18. Projected density of states (PDOS) on the Sn (blue) and Te (red) atoms in the (a) pure and (b) Sn-vacancy-containing SnTe (001) surfaces. The right panel shows the top views of the surfaces. The red dashed circle indicates the Sn-vacancy position. Black dashed lines represent the Fermi levels.

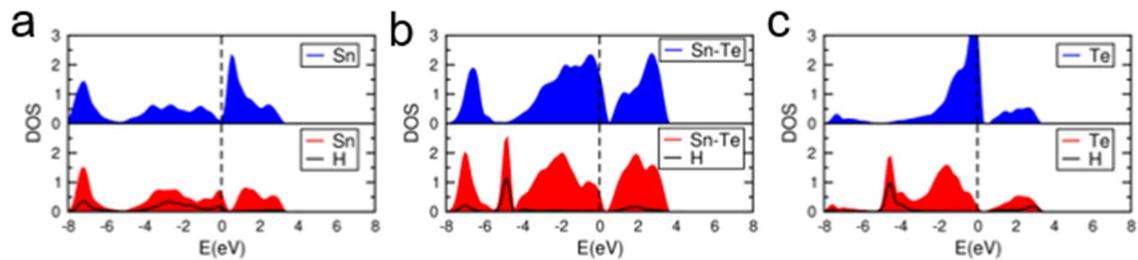

Figure S19. PDOS on hydrogen-hosting sites on (a) pure, (b) Sn-vacancy containing, and (c) partially oxidized SnTe (001) surfaces before (blue) and after (red) hydrogen adsorption. The solid black curve is the PDOS on the H atom. Black dashed lines represent the Fermi levels.

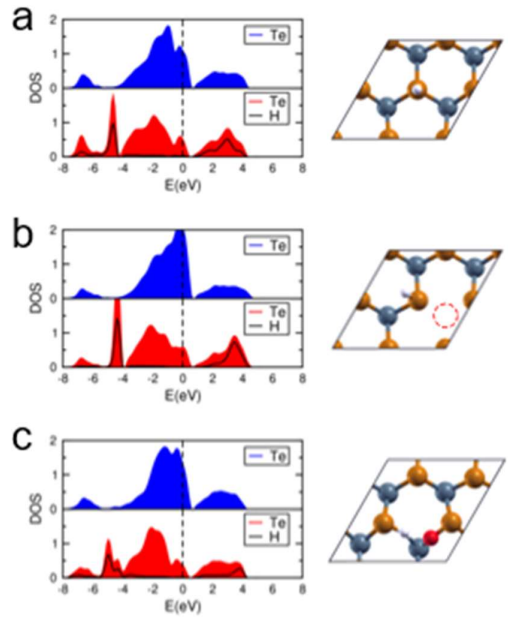

Figure S20. PDOS on the hydrogen-hosting Te atom on (a) pure, (b) Sn-vacancy-containing, and (c) partially oxidized SnTe (111) surface before (blue) and after (red) hydrogen adsorption. The right panel shows the top views of surfaces. The red dashed circle indicates the Sn-vacancy position. The solid black curve is the PDOS on the H atom. Black dashed lines represent the Fermi levels.

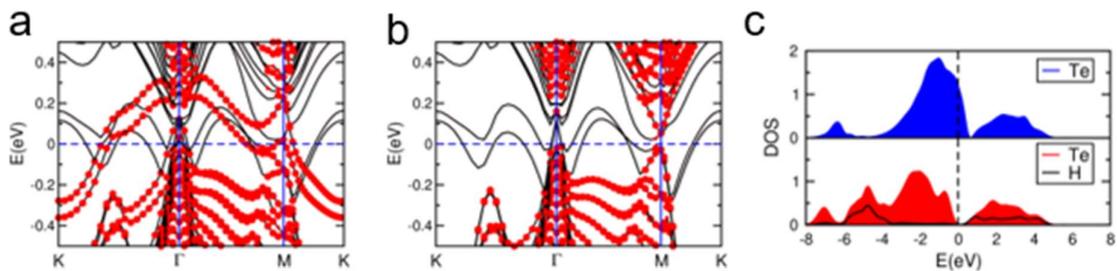

Figure S21. Band structures of $(1 \times 1)$ 48ML SnTe (111) slab (a) before and (b) after hydrogen adsorption. The sizes of red dots represent the contributions from (111) upper 6ML surface. (c) PDOS on Te atom before (blue) and after (red) hydrogen adsorption with 1ML coverage. The solid black curve is the PDOS on the H atom. The Fermi levels are set to zero.

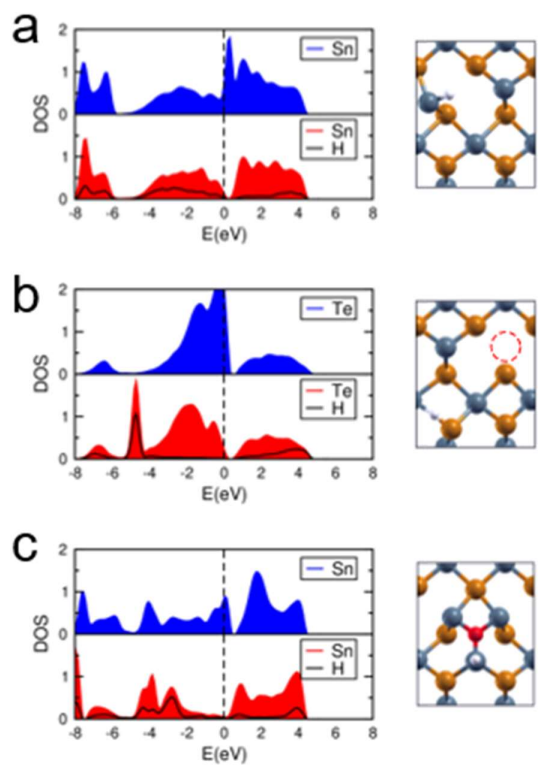

Figure S22. PDOS on the Te and Sn atoms in (a) pure and (b) Sn-vacancy-containing and (c) partially oxidized SnTe (211) surfaces before (blue) and after (red) hydrogen adsorption. The right panel shows the top views of atomic structures. The red dashed circle labels the position of a Sn vacancy. The solid black curve is the PDOS on the H atom. Black dashed lines represent the Fermi levels.

Supplementary note 2

**Hydrogen adsorption on Sn-terminated (111) surface**

Besides the Te-terminated (111) surfaces discussed in the main text, we also considered the HER activity of Sn-terminated (111) surfaces. As indicated by previous studies [19-21], surface reconstruction may occur on Sn-terminated (111) surfaces to cancel the internal dipole moment. Using first-principles DFT calculations, Wang et al. demonstrated that the (2×1)-1 Sn (111) surface with 50% Sn coverage could exist under a Sn-rich condition [19]. Figure S23 shows the band structure of a (2×1)-1Sn (111) slab whose thickness is 39ML. The mirror-symmetry protected Dirac cone is located near the Γ point, which has a small density of states at the Fermi level (Figure S24a). The hydrogen atom is repelled by the surface and $\Delta G_H$ is 0.85 eV (Figure S25).

Next we considered the effect of partial oxidation on the (2×1)-1Sn (111) surface by adding one oxygen atom onto the surface (Figure S26). The $\Delta G_H$ value largely decreases to 0.18 eV, indicating the high electrocatalytic activity for the HER. Electronic structure calculations (Figure S24b) show that the Dirac cone upshifts above the Fermi level, making the slab a p-type material. The charge transfer between the H atom and the surface enables the formation of a H-Te bond, which enhances the HER performance. These results regarding the magnitude of $\Delta G_H$ and the location of the Dirac cone are consistent with our finding on the (001) and Te-terminated (111) surface.

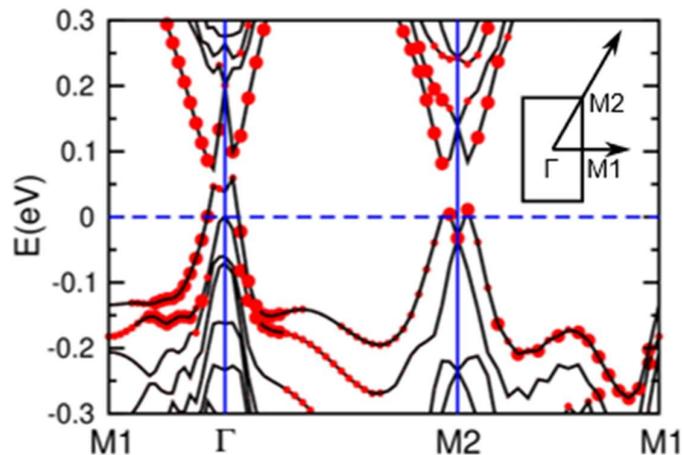

Figure S23. Band structure of pure $(2 \times 1)$-1Sn (111) surface with 39 ML thickness. The size of red dots represents the contributions from the upper 6ML in the (111) surface slab. The Fermi level is set to zero.

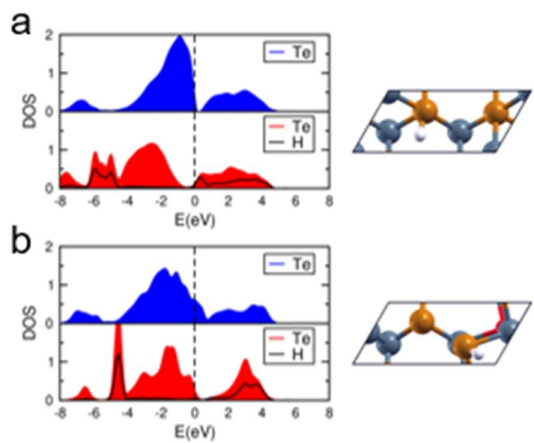

Figure S24. PDOS on hydrogen-hosting Te atom on (a) pure and (b) partially oxidized $(2 \times 1)$-1Sn (111) surfaces before (blue) and after (red) hydrogen adsorption. The right panel shows the top views of the atomic structures. The solid black curve is the PDOS on the H atom. Black dashed lines represent the Fermi levels.

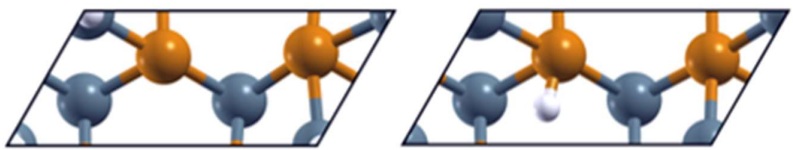

Figure S25. Top views of hydrogen adsorption on (2 × 1)-1Sn (111) surface at different adsorption sites.

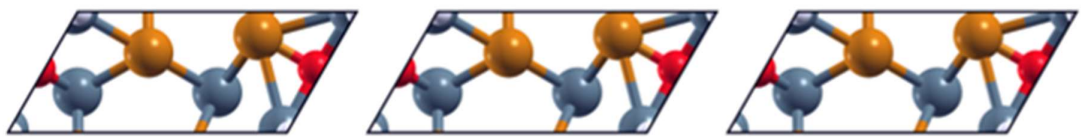

Figure S26. Top views of hydrogen adsorption on partially oxidized (2 × 1)-1Sn (111) surface at different adsorption sites.

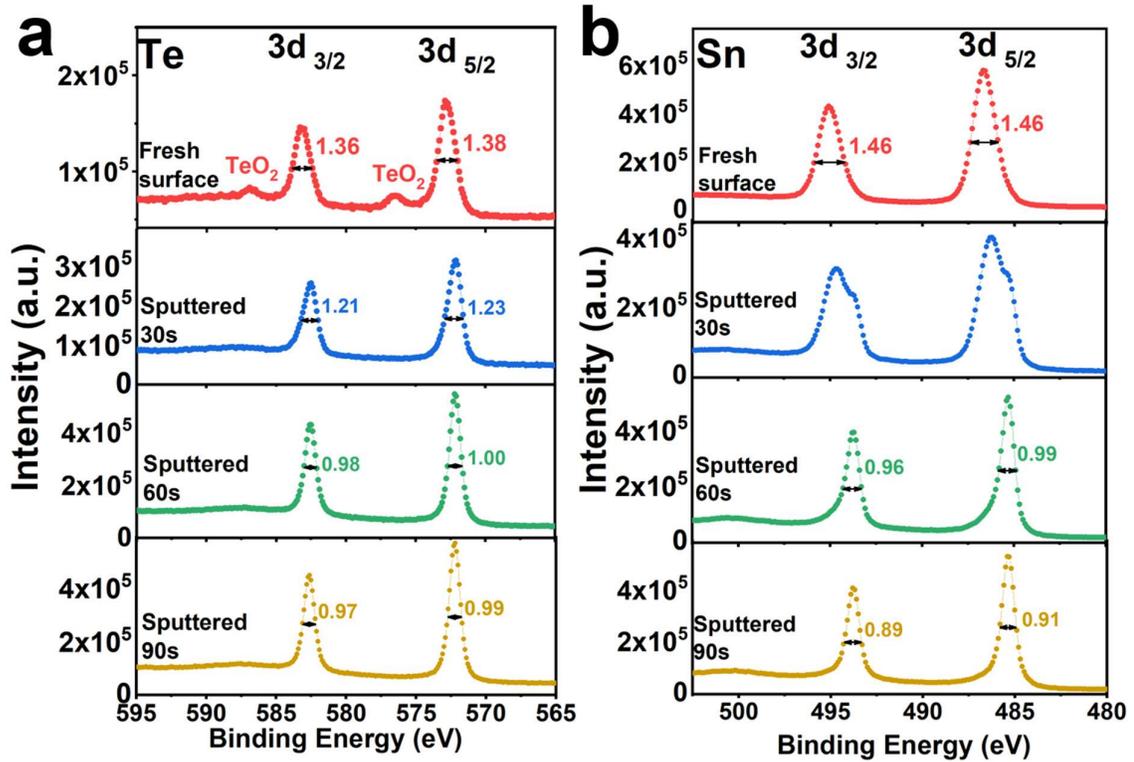

Figure S27. Raw data of Te 3d (a) and Sn 3d (b) XPS spectra of the SnTe (001) sample for its fresh surface, and those after sputtering for 30s, 60s, and 90s.

Supplementary note 3

**Theoretical Calculations**

For hydrogen adsorption simulations, we used $(1 \times 1)$, $c(2 \times 2)$, and $(2 \times 2)$ unit cells of SnTe (001) with 8ML thick; $(1 \times 1)$, $(\sqrt{3} \times \sqrt{3})$R30°, $(2 \times 2)$ and $(3 \times 3)$ unit cells of Te-terminated (111) with 10ML thick; $(2 \times 1)$-1Sn (111) with 10ML thick; $(1 \times 1)$ and $(2 \times 1)$ unit cells of SnTe (211) with 15ML thick. For electronic structure calculations, we used a 12ML SnTe (001) slab, a 24ML Te-terminated SnTe (111) slab, a 39ML $(2 \times 1)$-1Sn (111) slab, and a 24ML SnTe (211) slab. We used a $6 \times 6 \times 1$ Monkhorst-Pack k-point mesh for the structural optimization of the $(1 \times 1)$ SnTe (001) and (111) unit cells, and a $8 \times 8 \times 1$ k-point mesh for electronic structure calculations. We used a $6 \times 2 \times 1$ k-point mesh for the structural optimization of a $(1 \times 1)$ SnTe (211) unit cell and a $8 \times 4 \times 1$ k-point mesh for electronic structure calculations.

We calculated $\Delta G_H$ as:

$$\Delta G_H = \Delta E_H + \Delta E_{ZPE} - T\Delta S_H \qquad (1)$$

$\Delta E_H$ denotes the hydrogen absorption energy calculated by:

$$\Delta E_H = E[\text{SnTe} + \text{H}^{ad}] - E[\text{SnTe}] - \frac{1}{2}E[\text{H}_2] \qquad (2)$$

where $E[\text{SnTe} + \text{H}^{ad}]$ is the total energy of the SnTe slab adsorbed by a H atom, $E[\text{SnTe}]$ is the energy of the SnTe slab after removing the H atom, and $E[\text{H}_2]$ is the energy of a hydrogen molecule in the gas phase. We calculated the zero-point energy change of a H atom, $\Delta E_{ZPE}$, as

$$\Delta E_{ZPE} = E_{ZPE}[\text{H}^{ad}] - \frac{1}{2}E_{ZPE}[\text{H}_2] \qquad (3)$$

where $E_{ZPE}[\text{H}^{ad}]$ is the zero-point energy of the H atom adsorbed on the SnTe surface, and $E_{ZPE}[\text{H}_2]$ is the zero-point energy of a hydrogen molecule in the gas phase. $\Delta S_H$ is calculated as $-\frac{1}{2}S^0_{\text{H}_2}$, where $S^0_{\text{H}_2}$ is the entropy of $\text{H}_2$ in the gas phase at the standard condition (130.68 J $\cdot$ mol$^{-1}$ $\cdot$ K$^{-1}$ at $T = 298$ K and $p = 1$ bar)[22]. After substituting the last two terms into equation (3), we have $\Delta G_H = \Delta E_H + 0.25$ eV.

Table S1. Composition of the Bi$_2$Te$_3$:Fe layer of the SnTe (111) sample determined by EDS.

| Layer | Bi | Te | Fe |
|---|---|---|---|
| Bi$_2$Te$_3$:Fe | 41.1% | 58.1% | 0.8% |

Table S2. Free energies of hydrogen adsorption ($\Delta G_H$) on a Sn atom, a Te atom, and a Sn-Te bridge on pure and Sn-vacancy-containing surfaces of a ($2 \times 2$) SnTe (001) slab.

| | $\Delta G_H (eV)$ | |
|---|---|---|
| active site | Pure | Sn-vacancy-containing |
| Sn atom | 0.627 | 0.374 |
| Te atom | 1.138 | 0.43 |
| Sn-Te bridge | 1.139 | 0.304 |

Table S3. Löwdin charges of the hydrogen atom adsorbed on pure, Sn-vacancy-containing, and partially oxidized surfaces of SnTe (001), (111), and (211) slabs.

| | Löwdin Charge on Hydrogen | | |
|---|---|---|---|
| surface | (001) | (111) | (211) |
| pure | 1.1295 | 0.9903 | 1.1206 |
| Sn-vac | 1.0475 | 0.988 | 1.0445 |
| Oxi | 0.9962 | 0.9839 | 1.1136 |